# Nonparametric mixed logit model with market-level parameters estimated from market share data

Xiyuan Ren[1], Joseph Y. J. Chow[1,*], Prateek Bansal[2]

1. C2SMARTER University Transportation Center, New York University Tandon School of Engineering, Brooklyn, USA

2. Department of Civil and Environmental Engineering, College of Design and Engineering at National University of Singapore, Singapore

* Corresponding author: joseph.chow@nyu.edu

## ABSTRACT

We propose a nonparametric mixed logit model that is estimated using market-level choice share data. The model treats each market as an agent and represents taste heterogeneity through market-specific parameters by solving a multiagent inverse utility maximization problem, addressing the limitations of existing market-level choice models with parametric estimation. A simulation study is conducted to evaluate the performance of our model in terms of estimation time, estimation accuracy, and out-of-sample predictive accuracy. In a real data application, we estimate the travel mode choice of 53.55 million trips made by 19.53 million residents in New York State. These trips are aggregated based on population segments and census block group-level origin-destination (OD) pairs, resulting in 120,740 markets. We benchmark our model against multinomial logit (MNL), nested logit (NL), inverse product differentiation logit (IPDL), and the BLP models. The results show that the proposed model improves the out-of-sample accuracy from 65.30% to 81.78%, with a computation time less than one-tenth of that taken to estimate the BLP model. The price elasticities and diversion ratios retrieved from our model and benchmark models exhibit similar substitution patterns. Moreover, the market-level parameters estimated by our model provide additional insights and facilitate their seamless integration into supply-side optimization models for transportation design. By measuring the compensating variation for the driving mode, we found that a $9 congestion toll would impact roughly 60% of the total travelers. As an application of supply-demand integration, we showed that a 50% discount of transit fare could bring a maximum ridership increase of 9,402 trips per day under a budget of $50,000 per day.

**Keywords:** mixed logit, nonparametric estimation, market-level data, taste heterogeneity, mode choice.

## 1. Introduction

Discrete choice models (DCMs) play a central role in demand analysis and are widely applied to the field of transportation. Typically, the decision-making process of travel behavior is captured by a set of taste parameters reflecting the value that travelers place on time, cost, convenience, and other attributes of trip options (Bierlaire & Frejinger, 2008; Bowman & Ben-Akiva, 2001; McFadden, 1977). These parameters are usually estimated under the random utility maximization (RUM) theory and reveal considerable heterogeneity among different regions and population segments (Lai & Bierlaire, 2015; Reck et al., 2022; Train, 2016). Understanding the heterogeneity in taste parameters is crucial to forecasting travel demand (El Zarwi et al., 2017), designing mobility services (Parady et al., 2021), and implementing policies to improve efficiency, user satisfaction, and equity in transportation systems (Bills et al., 2022; Quddus et al., 2019; Yin & Cherchi, 2024).

Over the past decade, the availability of information and communication technology (ICT) data makes it possible to capture travel preferences of various population segments on a regional scale (Patwary & Khattak, 2022; Wan et al., 2021). Meanwhile, significant progress made in DCMs provides a powerful framework to estimate heterogeneous taste parameters (McFadden & Train, 2000; Sarrias, 2020; Schneider, 2013). Based on the data granularity and the distribution on parameters, existing DCMs considering taste heterogeneity can be categorized into three groups: (1) individual-level models with parametric distributions, (2) individual-level models with nonparametric (or semi-parametric) distributions, and (3) market-level models with parametric distributions.

McFadden and Train (2000) showed that any discrete choice model can be approximated by a mixed logit (MXL) from, providing a flexible distribution. Earlier studies applied MXL assuming different parametric distributions (e.g., normal, log-normal, or uniform distribution) on utility parameters. Given that the restrictions on parametric distribution might lead to bias in fitting the observed data, a number of individual-level nonparametric MXL models has been developed, including logit-mixed logit (LML) (Train, 2016), individual parameter logit (IPL) (Swait, 2023), and agent-based mixed logit (AMXL) (Ren & Chow, 2022). These models capture individual taste heterogeneity in a more flexible manner.

As for market-level models, Berry et al. (1995) presented a mixed logit form (also called BLP model), which specifies flexible substitution patterns through parametric assumption on consumer heterogeneity. Fosgerau et al. (2024) developed inverse product differentiation logit (IPDL) that is computationally efficient and generates similar substitution patterns as of BLP through non-hierarchical product segmentation[1] in multiple overlapping dimensions. The reviewed literature by Fosgerau et al. (2024) indicates a lack of market-level non-parametric models. Such flexible models could be crucial in situations when the number of products is relatively small while the tastes among local markets are extremely heterogeneous because even flexible product-segmentation would not be able to capture market-level heterogeneity. Taking travel mode choice as an example, a fast yet expensive service (e.g., ride-hailing) could be more attractive to urban employees, while a more affordable service with longer waiting time (e.g., public transit) could be preferred by seniors living in rural areas. To effectively design and manage transportation systems, it is crucial to consider population groups in different regions as separate markets and retrieve their diverse preferences from available datasets. Berry and Haile (2014) presented identification results for a nonparametric model of differentiated product markets. Nevertheless, they did not provide any real case studies, estimation tools, or benchmark comparisons.

[1] Non-hierarchical product segmentation is a method of dividing products into distinct groups based on similarities in attributes, without assuming any predefined hierarchical structure among the segments. In highly diverse markets where products can simultaneously belong to multiple categories, non-hierarchical segmentation more accurately reflects the true substitution and complementarity between products.

To address the gap, this study proposes a k-modal nonparametric mixed logit model in which (1) aggregate data of a large number of separate markets are used, with each market serving as an agent; (2) individual tastes are assumed to be homogeneous within a market while heterogeneous across markets; (3) an instrumental variable (IV) approach is used to handle endogeneity biases; and (4) a unique set of taste parameters is estimated for each market using a k-modal nonparametric approach under the RUM framework. The agent in our model does not need to represent a real market; instead, it can represent the aggregated choice made by any group of individuals sharing similar tastes. To this end, we call this model Group-Level Agent-based Mixed (GLAM) logit. Compared to existing mixed logit models, GLAM logit adopts a more restrictive mixture form, where taste heterogeneities are allowed only at the market level. The model is suitable for choice modeling tasks with the following characteristics: (1) only aggregate or market-level choice datasets are available, (2) taste heterogeneity within a market is minimal but substantial across markets, and (3) distributions of random parameters are nonparametric due to unobserved sociodemographic or spatial attributes.

As for the methodology, we first apply IV regression (Angrist & Krueger, 2001) to correct for endogeneity biases in market-level models. Then we formulate a multi-agent inverse utility maximization (MIUM) problem to estimate agent-level taste parameters. In each iteration, an inverse optimization (IO) problem (Ahuja & Orlin, 2001) is formulated for each agent to solve a set of parameters with constraints regarding inverted market share (Berry et al., 1995). A Method of Successive Averages (Sheffi & Powell, 1982) is applied to ensure iterative convergence and a stable fixed point. Finally, we add a classification step at the end of each iteration, which categorizes agent-specific parameters into several 'taste clusters' using a k-means algorithm (Hartigan & Wong, 1979). This allows for a k-modal distribution of taste parameters similar to what latent class models achieve (Greene & Hensher, 2003).

In the case study, we apply GLAM logit to mode choice modeling for the entire New York State (NYS). The experimental data is provided by Replica Inc., which contains 53.55 synthetic trips made by 19.53 million NYS residents on a typical Thursday in Fall 2019. We define the market based on geographic and socioeconomic attributes. Trips made by one of the four population segments (low-income, not-low-income, students, and seniors) and along a census block group-level origin-destination (OD) pair are segregated as a unique market, resulting in a total of 120,740 markets/agents. Six modes are considered, including driving, public transit, on-demand service, biking, walking, and carpool. We benchmark GLAM logit against state-of-the-art market-level DCMs. The parameters estimated by GLAM logit are further applied to analyze congestion pricing and optimize transit fare discounts. To facilitate future research, we upload the algorithms and examples to a GitHub repository: https://github.com/BUILTNYU/GLAM-Logit.

The remainder of the paper is organized as follows: in Section 2, we briefly review the existing research on individual-level and market-level DCMs and then list out our contributions. In Section 3, we introduce the mathematical formulation of the proposed GLAM logit model and its estimation algorithm. Section 4 sets up a simulation study to evaluate the performance of GLAM logit. Section 5 presents a concrete case study of New York statewide travel mode choice. Our key findings and future research directions are concluded in Section 6.

## 2. Literature review

### 2.1 Individual-level DCMs

Individual-level DCMs assume individuals make decisions by maximizing the overall utility they can expect to gain (Bowman & Ben-Akiva, 2001; McFadden, 1977). McFadden and Train (2000) proved that mixed random utility models such as mixed logit (MXL) provide a powerful framework to account for unobserved taste heterogeneity in DCMs. MXL is a mixture of multinomial logit (MNL) models with random parameters $\beta$ drawn from a probability distribution function $g(\beta|\theta)$, as shown in Eqs. (1)-(2).

$$U_{njt} = \beta^T X_{njt} + \varepsilon_{njt}\,, \quad \forall n \in N, \forall j \in J, t \in T \tag{1}$$

$$P_{nt}(j|X_{njt}, \beta) = \int \frac{e^{\beta^T X_{njt}}}{\sum_{j' \in J} e^{\beta^T X_{nj't}}}\, g(\beta|\theta) d\beta, \quad \forall n \in N, \forall j \in J, t \in T \tag{2}$$

where $N$ is the set of individuals; $J$ is the set of alternatives; $T$ is the set of choice situations. $U_{njt}$ is the overall utility of individual $n$ choosing alternative $j$ in situation $t$, which consists of a systematic utility $\beta^T X_{njt}$ and a random utility $\varepsilon_{njt}$ usually assumed to be independent and identically distributed (i.i.d.). $X_{njt}$ denotes a set of observed attributes of alternative $j$ for individual $n$ in situation $t$. $\beta$ is a vector of random taste parameters with a probability density $g(\beta|\theta)$, where $\theta$ represents the parameters of this distribution (e.g., mean and covariance matrix for normal distribution). The probability of individual $n$ choosing alternative $j$ in situation $t$ conditional on $X_{njt}$ and $\beta$ can be defined as Eq. (2).

MXL typically assumes that tastes vary across individuals, resulting in $\beta$ being indexed by $n$. When analyzing longitudinal choice data, MXL can be extended to capture unobserved inter- and intra-individual heterogeneity by allowing tastes to vary both across individuals and across choice situations encountered by the same individual. In that case, $\beta$ is indexed by $n$ and $t$, representing a more detailed form of mixture.

Though the mixed logit framework allows the distribution of taste parameters to be arbitrary, the mixing distribution is usually restricted to parametric distributions (e.g., normal, uniform, or triangular distribution), which might be problematic when taste heterogeneity deviates from the assumed parametric distribution (Hess, 2010; Sarrias, 2020). Alternatively, a number of studies proposed semi-parametric or nonparametric approaches to capture taste heterogeneity in a more flexible manner.

Fox et al. (2011) proposed a mixture estimator based on linear regression for recovering the joint distribution of taste heterogeneity in DCMs. The estimator is subjected to linear inequality constraints, and is computationally efficient compared to MXL models. Train (2016) proposed a logit-mixed logit (LML) model, in which the mixing distribution of parameters can be easily specified using splines, polynomials, step functions, and many other functional forms. Ren and Chow (2022) proposed an agent-based mixed logit (AMXL) model that is a variant of MXL model designed for ubiquitous data set. They used a hybrid machine learning/econometric approach to estimate deterministic, individual-specific parameters. Swait (2023) developed a nonparametric approach that combines an upper-level evolutionary algorithm and a lower-level gradient decent algorithm. The estimator directly maximizes the sample loglikelihood to obtain individual-level parameters.

Despite the several advancements, the studies mentioned above require individual-level data that are laborious to collect from surveys. Although information and communication technology (ICT) data can be ideal sources, their reliability is usually challenged for at least three reasons. First, since sensitive personal information has been removed due to privacy issues (Rao & Deebak, 2023), ICT data usually lack sufficient socioeconomic characteristics that are important in individual-level DCMs. Second, to construct individual choice datasets, data fusion approaches are required to get information about the attributes for all alternatives in the choice set (Krueger et al., 2023), which could bring additional uncertainties. Third, location-based data collection tends to have noises and measurement errors, especially when we directly use the information of individual geolocation (Ren et al., 2022). To this end, market-level models using aggregate data are still useful even though more and more data sources contain individual mobility profiles (He et al., 2020). It remains unclear how to transform flexible and non-parametric individual-level demand models to market level.

### 2.2 Market-level DCMs

Berry (1994) proposed an aggregate model for differentiated products under the random utility maximization (RUM) framework. The general idea of this model consists of two steps: in the first step, the model gets the mean utility across individuals by inverting the market share function; in the second step, the model estimates the relationship between product attributes and mean utility levels. The utility for product $j$ in market $t$ ($U_{jt}$) is defined as Eq. (3).

$$U_{jt} = \delta_{jt} + \varepsilon_{jt} = x_j\beta - \alpha p_{jt} + \xi_{jt} + \varepsilon_{jt}, \quad \forall j \in J, \forall t \in T \tag{3}$$

where $t$ is the index of markets (instead of choice situations in individual-level DCMs), $T$ is the set of markets, $\delta_{jt} = x_j\beta - \alpha p_{jt} + \xi_{jt}$ is the mean utility level, $x_j$ is a vector of $K$ attributes of product $j$, $\beta$ a set of parameters for these attributes, $p_{jt}$ is the price of product $j$ in the market $t$, $\alpha > 0$ is the parameter for price (also called marginal utility of the income), $\xi_{jt}$ represents the unobserved product attributes, and $\varepsilon_{jt}$ accounts for unobserved, market-specific randomness in preferences. Following Berry (1994) and Huo et al. (2024), the observed market shares ($s_{jt}$) and those predicted by the model are linked through invertible mapping in Eq. (4).

$$s_{jt} = f_j(\delta_t; \varphi) \;\rightarrow\; f_j^{-1}(s_t; \varphi) = \delta_{jt}\,, \quad \forall j \in J, \forall t \in T \tag{4}$$

where $f_j(.)$ is the demand function of product $j$, $\varphi$ is a set of parameters for the distribution of unobserved consumer preferences.

In line with this framework, Berry et al. (1995) presented a mixed logit model for market-level datasets, which is also called BLP model. The model incorporates unobservable taste heterogeneity across individuals, indexed by $n \in N$, through random parameters in the utility function (Eq. (5)).

$$U_{njt} = \delta_{njt} + \varepsilon_{njt} = x_j\beta_n - \alpha_n p_{jt} + \xi_{jt} + \varepsilon_{njt}, \quad \forall n \in N, \forall j \in J, \forall t \in T \tag{5}$$

where $\delta_{njt}$ is the mean utility level, $\varepsilon_{njt}$ is an error term that is usually assumed to be independent and identically distributed (i.i.d.) Gumbel variates, $\beta_n$ and $\alpha_n$ are individual-specific parameters assumed to be distributed as Eq. (6).

$$\binom{\alpha_n}{\beta_n} = \binom{\alpha}{\beta} + \begin{bmatrix} 1 & \cdots & 0 \\ \vdots & \ddots & \vdots \\ 0 & \cdots & 1 \end{bmatrix}_{K+1} \cdot \begin{bmatrix} \vartheta_{n,1} \\ \vdots \\ \vartheta_{n,K+1} \end{bmatrix} \tag{6}$$

where $\vartheta_{n,k}$ denotes individual $n's$ specific preference on the $k^{th}$ product variable and follows a normal distribution, $\vartheta_{n,k} \sim N(0, \sigma_k^2)$. Therefore, the utility level can be written as Eqs. (7)-(9).

$$\delta_{njt} = \bar{\delta}_{jt} + \mu_{njt}, \quad \forall n \in N, \forall j \in J, \forall t \in T \tag{7}$$

$$\bar{\delta}_{jt} = x_j \beta - \alpha p_{jt} + \xi_{jt}, \quad \forall j \in J, \forall t \in T \tag{8}$$

$$\mu_{njt} = [-p_{jt}, x_j] . \begin{bmatrix} \vartheta_{n,1} \\ \vdots \\ \vartheta_{n,K+1} \end{bmatrix}, \quad \forall n \in N, \forall j \in J, \forall t \in T \tag{9}$$

where $\bar{\delta}_{jt}$ denotes the market-level utility, and $\mu_{njt}$ denotes the unobserved consumer preferences capturing inter-individual taste heterogeneity. To this end, the BLP model allows variation across both individuals and markets. The share of product $j$ in market $t$ can be written in Eq. (10).

$$s_{jt} = \int \frac{e^{(\bar{\delta}_{jt} + \mu_{njt})}}{\sum_{j' \in J} e^{(\bar{\delta}_{j't} + \mu_{nj't})}} dF(\alpha_n, \beta_n), \qquad \forall j \in J, \forall t \in T \tag{10}$$

where $F(\alpha_n, \beta_n)$ is the probability density of a multivariate normal distribution. Since Eq. (10) does not have a closed form, the BLP model is estimated using a two-step iterative process (Nevo, 2000), which is computationally cumbersome due to numerical approximation of integral in Eq. (10).

Fosgerau et al. (2024) proposed inverse product differentiation logit (IPDL) to address the limitations of non-hierarchical product segmentation (Cardell, 1997) and provide faster estimation. IPDL assumes that differentiated products are segmented by $D$ dimensions/ attributes, with each product belonging to only one group in each dimension. In that case, the inverse demand function $(f_j^{-1}(.))$ is specified as Eqs. (11)-(12).

$$f_j^{-1}(s_t; \varphi) = \ln G_j(s_t; \varphi) + c_t = \delta_{jt}, \qquad \forall j \in J, \forall t \in T \tag{11}$$

$$\ln G_j(s_t; \varphi) = \left(1 - \sum_{d=1}^{D} \rho_d\right) \ln(s_{jt}) + \sum_{d=1}^{D} \rho_d \ln\left(\sum_{j' \in J_d} s_{j't}\right), \quad \forall j \in J, \forall t \in T \tag{12}$$

where $c_t$ is a constant for market $t$, $\rho_d$ is the grouping parameter for dimension $d$, and $J_d$ is a set of products grouped by dimension $d$. The higher value of $\rho_d$ implies that products in the same group are more similar in dimension $d$ than other dimensions. To this end, taste heterogeneity among consumers is captured by $\varphi = \{\rho_1, \rho_2, \dots, \rho_D\}$. Since the main utility level of the outside good is assumed to be zero ($\delta_{0t} = 0$), we have $\ln(s_{0t}) + c_t = \delta_{0t} = 0 \rightarrow c_t = -\ln(s_{0t})$. Linking this to Eqs. (11)-(12) we obtain Eq. (13) that relates the inverse market share to product attributes and taste heterogeneity.

$$\ln\left(\frac{s_{jt}}{s_{0t}}\right) = x_j\beta - \alpha p_{jt} + \sum_{d=1}^{D} \rho_d \ln\left(\frac{s_{jt}}{\sum_{j' \in J_d} s_{j't}}\right) + \xi_{jt}\,, \quad \forall j \in J, \forall t \in T \tag{13}$$

IPDL is a general form of multinomial logit (MNL) and nested logit (NL) model. MNL is obtained when there is no product segmentation ($\rho_d = 0, \forall d \in \{1,2,\dots,D\}$). NL is obtained when there is only one dimension ($D = 1$). Moreover, IPDL can be estimated using the two-stage least squares algorithm that is efficient to solve with large sample sizes (Fosgerau et al., 2024).

To sum up, existing market-level models capture taste heterogeneity by either assuming parametric distribution (e.g., normal distribution in the BLP model) or allowing flexible product segmentation (e.g., $D$-dimension product segmentation in IPDL). However, these parametric approaches could result in biased parameter estimation and inaccurate demand prediction, especially when individual tastes deviate from the parametric assumptions due to unobserved spatial or sociodemographic attributes (Farias et al., 2013; Ren & Chow, 2022). The only nonparametric market-level model we found is in Berry and Haile (2014)'s work, but they did not provide any real case or estimation tools, and they did not benchmark it against existing models. Moreover, parameters estimated by these models do not have one-on-one mapping with markets. A lack of market-specific parameters makes it hard to incorporate taste heterogeneity into system design models to link the demand and supply sides (Ren et al., 2024, Gómez-Lobo et al., 2022, Paneque et al., 2021).

### 2.3 Our contributions

The limitations mentioned above can be addressed if a group of homogeneous individuals or consumers is treated as a market, and a unique parameter can be estimated for each market. The proposed group-level agent-based mixed (GLAM) logit model, a k-modal non-parametric approach, achieves the same within the RUM framework.

Unlike individual-level MXL models, where tastes are allowed to vary across individuals and choice situations, and the BLP model, where tastes are allowed to vary across both individuals and markets, the GLAM logit model assumes that each market is characterized by a unique set of taste parameters ($\{\beta_t, \alpha_t, \xi_{jt}\}$). Within each market, individuals are treated as homogeneous. This means GLAM logit adopts a more restrictive mixture form, where variation is allowed only at the market level. This restriction is motivated by three key aspects. First, incorporating heterogeneity among individuals within markets (or choice situations encountered by the same individual) significantly increases computational complexity, making the model hard to estimate using the market-level data. Second, estimating market-level heterogeneity is sufficient for demand-supply integration because accounting for individual-level heterogeneity is anyway challenging in system-level supply-side optimization problems due to computational issues and data availability constraints. Third, with large ICT datasets, individuals can be aggregated based on geolocation and socio-demographic attributes, resulting in intra-market homogeneity and inter-market heterogeneity, which aligns well with our assumptions. To validate these points, a performance comparison of GLAM logit with existing market-level models is presented in Section 5.

To this end, the GLAM logit model makes sense when: (1) individual-level choice data is unavailable or unreliable to build individual models, (2) taste heterogeneities is difficult to specify using parametric distributions to capture variations in unobserved sociodemographic or spatial attributes, and (3) individual tastes are homogeneous within a market/agent while heterogeneous among markets/agents. These settings are realistic in the cases of travel destination choice, mode choice, or route choice modeling with large-scale datasets (He et al., 2020).

The significance of GLAM logit is as follows. First, it uses aggregate data which can be directly retrieved from available datasets. This avoids additional data fusion steps that introduce uncertainties and make it easier to address endogeneity biases. Second, it is the first practice-ready market-level non-parametric model. The agent-based logic and k-modal estimation allows modelers to capture taste heterogeneity by identifying an empirical distribution that fits to the observed data. Third, since each market's representative utility function is fully specified, market-specific parameters estimated by GLAM logit enable its efficient integration into optimization models to link the demand and supply sides for system design.

## 3. Proposed model

The proposed model is a k-modal nonparametric mixed logit model with agent-specific parameters estimated from market-level data. Notations used in this section are shown in Table 1.

**Table 1**
Notations used in the proposed model

| | |
|---|---|
| $U_{njt}$ | The total utility of individual $n$ choosing product $j$ in market $t$ |
| $U_{jt}$ | The total utility of product $j$ in market $t$ |
| $x_{jt}$ | The attributes of product $j$ in market $t$ |
| $p_{jt}$ | The price of product $j$ in market $t$ |
| $\beta_t$ | The parameters for product attributes in market $t$ |
| $\alpha_t$ | The parameter for product price in market $t$ |
| $\xi_{jt}$ | The unobserved attributes of product $j$ in market $t$ |
| $\bar{\delta}_{jt}$ | The general utility of product $j$ in market $t$ |
| $\mu_{njt}$ | The individual-specific unobserved utility of individual $n$ for product $j$ in market $t$ |
| $\varepsilon_{njt}$ | The unobserved error term in utility of individual $n$ choosing product $j$ in market $t$ |
| $\varepsilon_{jt}$ | The unobserved error term in utility of product $j$ in market $t$ |
| $s_{jt}$ | The market share of product $j$ in market $t$ |
| $c_{jt}$ | The exogenous attributes of product $j$ in market $t$ |
| $m_{jt}$ | The instrumental variable for the endogenous variable $p_{jt}$ |
| $z_{jt}$ | The explanatory variables in instrumental variable regression |
| $\gamma_j$ | The parameters in instrumental variable regression |
| $\tau_{jt}$ | The error term in instrumental variable regression |
| $\phi_t$ | The parameters in the control function |
| $\tilde{\varepsilon}_{jt}$ | The term in control function equation that is uncorrelated with $p_{jt}$ |
| $X_{jt}$ | The vector of all variables related to product $j$ for agent $t$ in GLAM logit |
| $\theta_t$ | The vector of all parameters for agent $t$ in GLAM logit |
| $\theta_0^m$ | The fixed-point prior of the $m^{th}$ taste cluster |
| $tol$ | The hyperparameter that ensures goodness-of-fit |
| $\omega_{tm}$ | The binary allocation variable indicating whether agent $t$ belongs to cluster $m$ |
| $lb, ub$ | The lower and upper boundaries for parameter estimation |
| $N$ | The set of all individuals |
| $N_t$ | The set of individuals in agent/market $t$ |
| $J$ | The set of all products |
| $T$ | The set of all agents/markets |

## 3.1 Architecture of GLAM logit

*3.1.1 Utility function and predicted market share*

Let us start from the utility function specified in the BLP model (Berry et al., 1995). Using Eqs. (5) and (7), the utility of individual $n$ in market $t$ choosing product $j$ ($U_{njt}$) can be written as Eq. (14).

$$U_{njt} = \bar{\delta}_{jt} + \mu_{njt} + \varepsilon_{njt}, \quad \forall n \in N, \forall j \in J, \forall t \in T \tag{14}$$

where, $\bar{\delta}_{jt} = x_{jt}\beta - \alpha p_{jt} + \xi_{jt}$ is the generic utility of product $j$ in market $t$, and $\mu_{njt}$ denotes the individual-specific unobserved utility. Since we assume individuals within a market are homogeneous, we have $\mu_{njt} = 0$ and $\varepsilon_{njt}$ following an i.i.d. Gumbel variate across $n$ for good $j$ in market $t$. Since tastes are considered heterogeneous across markets, $\alpha$, $\beta$, and $x_j$ are indexed by $t \in T$. Therefore, in GLAM logit the utility function can be written as Eq. (15).

$$U_{njt} = U_{jt} = \bar{\delta}_{jt} + \varepsilon_{jt} = x_{jt}\beta_t - \alpha_t p_{jt} + \xi_{jt} + \varepsilon_{jt}, \quad \forall n \in N_t, \forall j \in J, \forall t \in T \tag{15}$$

where $N_t$ is the subset of individuals belonging to market $t$ and $\varepsilon_{jt}$ follows an i.i.d. Gumbel variate. Hence, the market share of product $j$ in market $t$ is predicted as Eq. (16), and the logarithm form of a ratio between two market shares can be presented as Eq. (17).

$$s_{jt} = \frac{e^{\bar{\delta}_{jt}}}{\sum_{j' \in J} e^{\bar{\delta}_{j't}}}, \quad \forall j \in J, \forall t \in T \tag{16}$$

$$\ln\left(\frac{s_{jt}}{s_{j't}}\right) = \ln\left(\frac{e^{\bar{\delta}_{jt}}}{e^{\bar{\delta}_{j't}}}\right) = \bar{\delta}_{jt} - \bar{\delta}_{j't} \quad \forall j, j' \in J, j \neq j', \forall t \in T \tag{17}$$

where if we assume there is an outside good with a systematic utility equal to zero, we obtain $\ln\left(\frac{s_{jt}}{s_{0t}}\right) = \bar{\delta}_{jt} = x_{jt}\beta_t - \alpha_t p_{jt} + \xi_{jt}$, which aligns with the original work of Berry (1994).

*3.1.2 Endogeneity correction with instrumental variables*

In our case, endogeneity bias could arise from the correlation between product price ($p_{jt}$) and the error term in the random utility ($\varepsilon_{jt}$), i.e., $Cov(p_{jt}, \varepsilon_{jt}) \neq 0$. To address the bias, we adopt the instrumental variable approach (Angrist & Krueger, 2001). In the first stage, the endogenous variable ($p_{jt}$) is regressed on a set of instrumental variables ($m_{jt}$) and exogenous variables ($c_{jt}$), as shown in Eq. (18).

$$p_{jt} = \hat{p}_{jt} + \tau_{jt} = z_{jt}^T \gamma_j + \tau_{jt}, \quad \forall j \in J, \forall t \in T \tag{18}$$

where $z_{jt} = \{m_{jt}, c_{jt}\}$ is a vector of independent variables. $\gamma_j$ is a set of parameters to be estimated. $\hat{p}_{jt} = z_{jt}^T \gamma_j$ is the predicted price in the instrumental regression. The error term $\tau_{jt}$ captures the influence of unobserved attributes that impact $p_{jt}$ but are not included in $z_{jt}$. The instrumental variables are constructed to ensure that the correlation between the predicted price and the error term in the random utility is zero, i.e., $Cov(\hat{p}_{jt}, \varepsilon_{jt}) = 0$.

In the second stage, the choice model is estimated with the utility function after endogeneity correction, as shown in Eq. (19).

$$U_{jt} = \bar{\delta}_{jt} + \varepsilon_{jt} = x_{jt}\beta_t - \alpha_t \hat{p}_{jt} + \xi_{jt} + \varepsilon_{jt}, \quad \forall j \in J, \forall t \in T \tag{19}$$

where $x_{jt}, \hat{p}_{jt}$ are explanatory variables for product $j$ in market $t$, and $\beta_t, \alpha_t, \xi_{jt}$ are market-level parameters to be estimated. Since $\hat{p}_{jt}$ is uncorrelated with $\varepsilon_{jt}$, an unbiased estimate of taste parameter for price ($\alpha_t$) can be obtained. If we use a compact form, the utility function and the logarithm form of the market share ratio can be rewritten as Eqs. (20)-(21).

$$U_{jt} = V(X_{jt}, \theta_t) + \varepsilon_{jt} = \theta_t^T X_{jt} + \varepsilon_{jt}, \quad \forall j \in J, \forall t \in T \tag{20}$$

$$\ln\left(\frac{s_{jt}}{s_{j't}}\right) = \ln\left(\frac{e^{V(X_{jt},\theta_t)}}{e^{V(X_{j't},\theta_t)}}\right) = \theta_t^T (X_{jt} - X_{j't}), \quad \forall j, j' \in J, j \neq j', \forall t \in T \tag{21}$$

where $V(X_{jt}, \theta_t)$ is a function of systematic utility, $X_{jt} = \{x_{jt}, p_{jt}, 1\}$ is a vector of attributes, and $\theta_t = \{\alpha_t, -\beta_t, \xi_{jt}\}$ is a vector of parameters. Using the two-stage estimation approach, standard errors of the second stage need to be computed via bootstrapping (Krueger et al., 2023).

### 3.2 K-modal nonparametric estimation algorithm for GLAM logit

#### *3.2.1 Multiagent inverse utility maximization (MIUM) problem for model estimation*

Following the work of Xu et al. (2018) and Ren and Chow (2022), we propose a multiagent inverse utility maximization (MIUM) problem to estimate the GLAM logit model. The agent-level parameters ($\theta_t$) can be jointly and nonparametrically estimated by solving a MIUM problem under $L_2$-norm as a convex quadratic programming (QP) problem. Considering that the empirical distribution of taste parameters can be multimodal, we use *M* fixed-point priors referring to *M* peaks in the multimodal distribution. Similar to the latent class logit model, this would allow modelers to identify taste clusters (Greene & Hensher, 2003). The formulation of a MIUM problem is shown in Eqs. (22)-(29).

$$\min_{\theta_0^m, \theta_t} \sum_{m=1}^{M} \sum_{t \in T} \omega_{tm} (\theta_0^m - \theta_t)^2 \tag{22}$$

subject to:

$$\theta_t^T (X_{jt} - X_{j't}) \geq \ln\left(\frac{s_{jt}}{s_{j't}}\right) - tol, \quad \forall j, j' \in J, j \neq j', \forall t \in T \tag{23}$$

$$\theta_t^T (X_{jt} - X_{j't}) \leq \ln\left(\frac{s_{jt}}{s_{j't}}\right) + tol, \quad \forall j, j' \in J, j \neq j', \forall t \in T \tag{24}$$

$$\theta_t \geq lb, \quad \forall t \in T \tag{25}$$

$$\theta_t \leq ub, \quad \forall t \in T \tag{26}$$

$$\theta_0^m = \frac{\sum_{t \in T} \theta_t \omega_{tm}}{\sum_{t \in T} \omega_{tm}}, \quad \forall m \in \{1, 2, \dots, M\} \tag{27}$$

$$\sum_{m=1}^{M} \omega_{tm} = 1, \qquad \forall t \in T \tag{28}$$

$$\omega_{tm} \in \{0,1\}, \qquad \forall t \in T, \forall m \in \{1,2,\dots,M\} \tag{29}$$

where $\theta_0^m$ is the $m^{th}$ fixed-point prior corresponding to a peak of the multimodal distribution; $\theta_t$ are agent-specific parameters; $T$ is the set of all markets (agents); $M$ is total number of peaks or taste clusters; $\omega_{tm}$ are introduced as binary allocation variables with $\omega_{tm} = 1$ indicating that parameters of agent $t$ belong to peak $m$. Eq. (22) defines the objective function, which is to minimize the squared distance between fixed-point priors and agent-level parameters. Eqs. (23)-(24) ensure that the predicted market share ratios are close to the observed market share ratios within a tolerance level $tol$, which is a manually set constant. A smaller value of $tol$ leads to higher goodness-of fit but might result in infeasible solutions. Eqs. (25)-(26) determine the parameter boundary for estimation, in which $lb$ and $ub$ specifies the lower and upper boundaries of parameters $\theta_t$. Eq. (27) ensures that the $m^{th}$ fixed-point prior comes from the mean value of agent parameters belonging to cluster $m$. Eq. (28) ensures that each agent belongs to only one cluster. Eq. (29) defines $\omega_{tm}$ as binary variables.

### *3.2.2 Proposed algorithm*

Solving the model in Eqs. (22)-(29) as a single QP would be computationally costly as it would lead to a highly sparse diagonal matrix and nonlinear constraints. Instead, we propose a decomposition method to initialize $\theta_0^m$ and $\omega_{tm}$ and update them iteration by iteration. In each iteration, we solve Eqs. (22)-(26) $|T|$ times with $\theta_t$ as the decision variables and $\omega_{tm}$, $\theta_0^m$ fixed, which results in much smaller QP problems. At the end of each iteration, we apply the k-means algorithm (Hartigan & Wong, 1979) to $\theta_t$ to identify $M$ taste clusters and update $\omega_{tm}$ using the classification results. The variable $\omega_{tm} = 1$ if agent $t$ is classified to cluster $m$, which satisfies the constraints in Eqs. (28)-(29). Finally, fixed-point priors $\theta_0^m$ are updated using Eq. (27).

We set a stopping criterion, where the percentage change of $\theta_0^m$ is smaller than a threshold $\epsilon$, to check if the algorithm has converged. If so, we output the estimated market-specific parameters $\theta_t$. Otherwise, we use the updated $\theta_0^m$ and $\omega_{tm}$ for the next iteration. The iterations continue until all priors $\theta_0^m$ stabilize. The subproblem with fixed $\omega_{tm}$, $\theta_0^m$ can be solved using any optimizer software or package that can handle QP like Gurobi, CVXPY, etc. The iterative updating method used in our study is the Method of Successive Averages (MSA), which ensures that the decomposition algorithm converges to a fixed point (Sheffi & Powell, 1982). The whole estimation approach is summarized in Algorithm 1.

---

**Algorithm 1. Parameter estimation in GLAM logit**

---

1. Given observed variables and market share $X_{jt}$, $s_{jt}$, set the iteration index $i$ to zero, initialize the fixed-point priors $\theta_0^{m(i)}$, and randomly assign market $t$ to one of the $M$ clusters.
2. For each $t \in T$, solve a QP problem if $\omega_{tm}^{(i)}$=1 to get $\theta_t^{(i)}$:
   $\min_{\theta_i^{(i)}} \delta_{ik}^{(i)} (\theta_0^{(i)} - \theta_i^{(i)})^2$ subject to constraints in Eqs. (23)-(26)

---

3. Apply the k-means algorithm to $\theta_t^{(i)}$ to identify $M$ taste clusters, and update to get $\omega_{tm}^{(i+1)}$ using classification results.
4. Set average to $y^{m(i)} = \frac{\sum_{t\in T} \theta_t^{(i)} \omega_{tm}^{(i+1)}}{\sum_{t\in T} \omega_{tm}^{(i+1)}}$, $\forall m \in \{1,2, \dots, M\}$ as shown in Eq. (27).
5. Using MSA to update and get $\theta_0^{m(i+1)}$:

$$\theta_0^{m(i+1)} = \frac{n}{n+1}\theta_0^{m(i)} + \frac{1}{n+1}y^{m(i)}, \qquad \forall m \in \{1,2, \dots, M\}$$

6. If the stopping criteria for $\theta_0^m$ reached, stop and output $\theta_0^{m(i)}$, $\theta_t^{(i)}$, $\omega_{tm}^{(i+1)}$; else, set $i = i + 1$ and go back to Step 2

The computational time is proportional to the total number of iterations and the time spent at each iteration. In each iteration, the MIUM problem is decomposed into $|T|$ QP problems. For each QP problem, the computational time is proportional to the number of constraints decided by the size of the choice set $|J|$. Hence, the computational time of our proposed algorithm would increase proportionally by ($|T| \times |J|$). However, the $|T|$ markets can be parallelized using a MapReduce approach since $\theta_0^m$ and $\omega_{tm}$ are fixed values in their QP subproblems. To this end, though GLAM logit takes a longer estimation time compared to IPDL, it is at least faster than BLP in which the estimation problem cannot be decomposed and solved in parallel. Moreover, the MSA algorithm can be further replaced with faster iterative algorithms like Method of Self-Regulated Average (MSRA) (Liu et al., 2009).

### 3.3 Out-of-sample prediction and hyperparameters in GLAM logit

#### *3.3.1 Out-of-sample prediction approach*

Since GLAM logit only specifies parameters for in-sample markets used for training, we apply a $\mathcal{K}$-nearest neighbors (KNN) approach (Cover & Hart, 1967) to retrieve taste parameters for out-of-sample markets. Algorithm 2 summarizes the prediction approach. It begins by defining a vector of attributes $x$ with $H$ dimensions to differentiate markets and computes the L2-norm $d(x_q, x_t)$ between a new market $q$ and all in-sample markets. Based on these distances, the $\mathcal{K}$-nearest neighbors are identified, denoted as $S_{\mathcal{K}}(x_q)$. The taste parameters for the new market, $\theta_q$, are then estimated as a weighted average of the parameters of its neighbors, with weights inversely proportional to the distances. Finally, $\theta_q$ is used to predict the market share of product $j$ in market $q$ using a multinomial logit formula. The process ensures accurate predictions for new markets by leveraging the correlations between market-level attributes and taste parameters.

**Algorithm 2. Out-of-sample prediction in GLAM logit**

1. Define a vector $x$ that includes $H$ attributes differentiating the markets. For a new market $q$, calculate the L2-norm between $x_q$ and all $x_t$ of in-sample markets. $d(x_q, x_t) = \sqrt{\sum_{h=1}^{H}(x_{q,h} - x_{t,h})^2}$.
2. Based on $d(x_q, x_t)$, select $\mathcal{K}$ nearest in-sample markets as the neighbors of the new market $q$: $S_{\mathcal{K}}(x_q) = \{m_{(1)}, m_{(2)}, \dots, m_{(\mathcal{K})}\}$, where $m_{(k)}$ denotes the index of the $k^{th}$ nearest neighbor.
3. Retrieve the taste parameters for the new market by computing a weighted average of its neighbors: $\theta_q = \frac{\sum_{t\in S_{\mathcal{K}}(x_q)} w_t \cdot \theta_t}{\sum_{t\in S_{\mathcal{K}}(x_q)} w_t}$, where $w_t = \frac{1}{d(x_q, x_t)}$.
4. Use $\theta_q$ for prediction. The market share of product $j$ in market $q$ is computed as: $s_{jq} = \frac{e^{\theta_q X_{jq}}}{\sum_{j' \in J} e^{\theta_q X_{j'q}}}$

*3.3.2 Hyperparameters in GLAM logit*

There are four hyperparameters in the proposed model: (1) the initial value of the fixed point priors ($\theta_0^{m(0)}$) in Algorithm 1. Starting with a good initial value can speed up the estimation algorithm; (2) the tolerance level ($tol$) in Eqs. (23)-(24). A small value of $tol$ leads to better goodness-of-fit but may result in infeasible solutions if the taste parameters cannot by tuned to make the predicted market share close to the observed one; (3) the number of tastes clusters ($M$) in Algorithm 1, which determines the shape of the nonparametric distribution estimated by GLAM logit. A small value of $M$ helps reduce the risk of overfitting but may result in poorer goodness-of-fit due to underfitting; and (4) the number of nearest neighbors ($\mathcal{K}$) in Algorithm 2. A large value of $\mathcal{K}$ is less sensitive to noise but may miss local patterns. These hyperparameters can be determined by drawing a balance between estimation time, estimation accuracy, and out-of-sample predictive accuracy. In Section 4, we use a simulation study to show how this works.

## 4. Simulation study

In this section, we present an extensive simulation evaluation of the GLAM logit model with different combinations of hyperparameters. The performance of GLAM logit is measured in terms of estimation time, estimation accuracy, and out-of-sample predictive accuracy. The simulation study provides insights into the sensitivity of GLAM logit to its hyperparameters. The Python codes for data generation, model estimation, and results analysis can be accessed through this link: https://github.com/BUILTNYU/GLAM-Logit/blob/main/simulation_study_v2.py.

### 4.1 Data generation

For the simulation study, we rely on synthetic choice data (Krueger et al., 2021), which we generate as follows: the choice sets comprise four unlabeled alternatives, which are characterized by three attributes. In each market, individuals are assumed to be homogeneous utility maximisers and to evaluate the alternative based on the utility function shown in Eq. (30), resulting in a market share shown in Eq. (31).

$$U_{jt} = X_{jt}\theta_t + \varepsilon_{jt}, \quad \forall j \in J, \forall t \in T \tag{30}$$

$$s_{jt} = \frac{e^{X_{jt}\theta_t}}{\sum_{j'\in J} e^{X_{j't}\theta_t}}, \quad \forall j \in J, \forall t \in T \tag{31}$$

where $j$ is the index of alternatives, $t$ is the index of markets, and $T$ is the set of markets. The taste parameters $\theta_t$ are drawn from a Gaussian distribution shown in Eqs. (32)-(33).

$$\theta_t | \zeta, \Sigma, \sim N(\zeta, \Sigma), \quad \forall t \in T \tag{32}$$

$$\Sigma = diag(\sigma)\, \Omega\, diag(\sigma) \tag{33}$$

where $\zeta$ is the mean vector, $\sigma$ is the standard deviation vector, $\Omega$ is the correlation matrix, and $\Sigma$ is the covariance matrix. We consider two scenarios, in which the distribution of taste parameters can be unimodal or multimodal. In the unimodal scenario ($\alpha = 1$), we set $\zeta = [-0.5, -0.5, 0.5]$, $\sigma = [1,1,1]$, and $\Omega = \begin{bmatrix} 1 & 0.5 & 0 \\ 0.5 & 1 & 0 \\ 0 & 0 & 1 \end{bmatrix}$, i.e. the total variance of each random parameter is twice the absolute value of its mean, and the first two taste parameters are correlated with each other. In the multimodal scenario ($\alpha = 3$), we mix three unimodal Gaussian distributions with $\zeta_1 = [2, 2, 3]$, $\zeta_2 = [-0.5, -0.5, 0.5]$, $\zeta_2 = [-3, -3, 2]$, $\sigma_1 = \sigma_2 = \sigma_3 = [1,1,1]$, and $\Omega_1 = \Omega_2 = \Omega_3 = \begin{bmatrix} 1 & 0.5 & 0 \\ 0.5 & 1 & 0 \\ 0 & 0 & 1 \end{bmatrix}$, i.e. the means of the three Gaussian distributions are spaced 2.5 apart from each other. In both scenarios, the alternative attributes $X_{jt}$ are drawn from $Uniform(0,5)$, which leads to an error rate of approximately 30%, i.e. 30% of the cases individuals in a market deviate from the systematically best alternative due to the stochastic utility component.

For out-of-sample prediction, we generate two additional variables ($lat, lon$) that are strongly correlated with the true taste parameters (Pearson correlation coefficients = 0.8). This implies that taste parameters for new markets can be inferred from in-sample markets with similar attributes, such as location. We generate a test set with a market size equal to 20% of the training sample. The mean values of $lat$ and $lon$ are set to zero. Their standard deviations are set to 10. These variables are used to run the K-nearest-neighbors algorithm for retrieving taste parameters to make predictions for out-of-sample markets.

We further let the total number of training markets $|T|$ take a value in {500, 5,000}. In the multimodal scenario, each unimodal distribution accounts for one-third of the total markets. For each scenario and for each value of $|T|$, we perform 20 replications. In each replication, $\theta_t$, $X_{jt}$ and ($lat, lon$) are generated using a unique random seed.

## 4.2 Accuracy assessment

We evaluate the performance of GLAM logit in terms of its ability to recover true taste parameters and its out-of-sample predictive accuracy.

### *4.2.1 Estimation accuracy*

To assess how well the proposed model perform at recovering parameters, we calculate the root mean square error (RMSE) for the mean vector $\zeta$ and the covariance matrix $\Sigma$. Given a collection of true value $\theta$ and its estimate $\hat{\theta}$, RMSE is defined as $RMSE(\hat{\theta}) = \sqrt{\frac{1}{H}(\hat{\theta} - \theta)^T(\hat{\theta} - \theta)}$, where $H$ denotes the total number of scalar parameters. For both unimodal and multimodal scenarios, we calculate the mean and covariance of the taste parameters across all markets, resulting in 3 values for $\zeta$ and $3 \times 3 = 9$ values for $\Sigma$.

*4.2.2 Predictive accuracy*

We construct three metrics to measure the prediction accuracy, including mean absolute error ($MAE$), overall accuracy ($OA$), and adjusted R-square ($ARS$), as shown in Eqs. (34)-(36).

$$MAE = \frac{1}{|J||T|} \sum_{t \in T} \sum_{j \in J} |\hat{s}_{jt} - s_{jt}| \tag{34}$$

$$OA = \frac{1}{|T|} \sum_{t \in T} \sum_{j \in J} \min\,(\hat{s}_{jt}, s_{jt}) \tag{35}$$

$$ARS = 1 - \frac{RSS(|T| - F)}{TSS(|T| - 1)} \tag{36}$$

where $\hat{s}_{jt}$ is the predicted market share, $s_{jt}$ is the true market share, $RSS$ is the residual sum of squares measured as $\sum_{t \in T} \sum_{j \in J} \left(s_{jt} - \hat{s}_{jt}\right)^2$, $TSS$ is the total sum of squares measured as $\sum_{t \in T} \sum_{j \in J} \left(s_{jt} - \bar{s}_{jt}\right)^2$, and $F$ is the total number of parameters. In general, $MAE$ measures the average prediction error per market share, $OA$ measures the percentage of market share that is correctly predicted, and $ARS$ is a common metric for the summary of regression models.

### 4.3 Experimental results

Using the simulated datasets, we compare the performance of GLAM logit models under different hyperparameter configurations. We set the stopping criteria $\epsilon$ to $10^{-3}$. All of the experiments were conducted on a local machine with Intel(R) Core(TM) i7-10875H CPU and 32GB installed RAM. The Gurobi package in Python was used to estimate GLAM logit.

We first check the sensitivity of GLAM logit to initial values ($\theta_0^{m(0)}$ or $ini$) and tolerance value ($tol$). For each experimental scenario defined by $\alpha$ and $|T|$, we set $tol$ to one of the values in {0.1, 0.5, 2.0} and consider two sets of initial values: [-0.5, -0.5, 0.5], which is close to the mean of true taste parameters, and [-2, -2, 2], which is farther away. For a fair comparison, the number of taste clusters ($M$) is set to 1. Table 2 compares the estimation performance of these models, revealing several interesting findings. First, while estimation time increases with the sample size ($|T|$), the estimation accuracy improves with the sample size. For instance, with $\alpha = 1$ and $tol = 0.1$, the mean RMSE value for $\zeta$ is 0.0067 for 500 samples and decreases to 0.0019 for 5,000 samples. This is likely because GLAM logit becomes more stable as the sample size increases to 5,000. Second, the choice of initial values does not significantly impact estimation accuracy. With $tol = 0.1$ and the same $\alpha$ and $|T|$, the RMSEs of $\zeta$ and $\Sigma$ for different initial values are exactly the same. This proves that GLAM logit achieves global convergence when the tolerance level is relatively small. However, bad initial values may increase the estimation time. In the unimodal scenario, with $|T| = 5{,}000$ and $tol = 0.1$, using [−0.5, −0.5, 0.5] requires 15.11 seconds, while [−2, −2, 2] increases the estimation time to 43.45 seconds. Finally, smaller values of $tol$ (e.g., 0.1) consistently yield lower RMSE values and shorter estimation times, probably because a smaller $tol$ results in fewer local optima. However, this does not mean that $tol$ should be as small as possible, as a small $tol$ could lead to infeasible solutions and overfitting. Based on our experiments, we recommend setting $tol$ in the range of 0.1 to 0.5, although this may vary across different cases.

**Table 2**
Estimation performance of GLAM logit under different initial values ($ini$) and tolerance levels ($tol$)

| $\alpha$ | \|T\| | Method | RMSE ($\zeta$) | | RMSE($\Sigma$) | | Iterations | | Time [s] | |
|---|---|---|---|---|---|---|---|---|---|---|
| | | | Mean | SE | Mean | SE | Mean | SE | Mean | SE |
| 1 | 500 | ini = [-0.5,-0.5,0.5], tol = 0.1 | 0.0067 | 0.0030 | 0.0606 | 0.0047 | 2.00 | 0.00 | 1.53 | 0.07 |
| | | ini = [-0.5,-0.5,0.5], tol = 0.5 | 0.0128 | 0.0055 | 0.1734 | 0.0095 | 2.60 | 0.49 | 2.00 | 0.39 |
| | | ini = [-0.5,-0.5,0.5], tol = 2.0 | 0.0252 | 0.0074 | 0.3649 | 0.0173 | 4.00 | 0.89 | 3.07 | 0.69 |
| | | ini = [-2,-2,2], tol = 0.1 | 0.0067 | 0.0030 | 0.0606 | 0.0047 | 6.20 | 0.40 | 4.70 | 0.30 |
| | | ini = [-2,-2,2], tol = 0.5 | 0.0130 | 0.0055 | 0.1735 | 0.0095 | 11.00 | 0.55 | 8.24 | 0.43 |
| | | ini = [-2,-2,2], tol = 2.0 | 0.0262 | 0.0076 | 0.3649 | 0.0173 | 18.95 | 0.80 | 13.67 | 0.76 |
| 1 | 5,000 | ini = [-0.5,-0.5,0.5], tol = 0.1 | 0.0019 | 0.0008 | 0.0598 | 0.0020 | 2.00 | 0.00 | 15.11 | 0.10 |
| | | ini = [-0.5,-0.5,0.5], tol = 0.5 | 0.0039 | 0.0014 | 0.1723 | 0.0027 | 2.00 | 0.00 | 15.02 | 0.09 |
| | | ini = [-0.5,-0.5,0.5], tol = 2.0 | 0.0068 | 0.0026 | 0.3640 | 0.0032 | 2.00 | 0.32 | 14.97 | 2.37 |
| | | ini = [-2,-2,2], tol = 0.1 | 0.0019 | 0.0008 | 0.0598 | 0.0020 | 6.00 | 0.00 | 43.45 | 1.03 |
| | | ini = [-2,-2,2], tol = 0.5 | 0.0041 | 0.0015 | 0.1723 | 0.0027 | 11.00 | 0.00 | 79.92 | 1.39 |
| | | ini = [-2,-2,2], tol = 2.0 | 0.0074 | 0.0033 | 0.3650 | 0.0032 | 18.90 | 0.30 | 130.56 | 8.14 |
| 3 | 500 | ini = [-0.5,-0.5,0.5], tol = 0.1 | 0.0077 | 0.0031 | 0.1155 | 0.0198 | 2.00 | 0.00 | 1.59 | 0.03 |
| | | ini = [-0.5,-0.5,0.5], tol = 0.5 | 0.0187 | 0.0066 | 0.3927 | 0.0357 | 2.50 | 0.50 | 1.95 | 0.38 |
| | | ini = [-0.5,-0.5,0.5], tol = 2.0 | 0.4420 | 0.0108 | 1.0099 | 0.0555 | 3.80 | 0.75 | 2.91 | 0.56 |
| | | ini = [-2,-2,2], tol = 0.1 | 0.0077 | 0.0031 | 0.1155 | 0.0198 | 5.05 | 0.59 | 3.86 | 0.49 |
| | | ini = [-2,-2,2], tol = 0.5 | 0.0186 | 0.0067 | 0.3927 | 0.0357 | 8.55 | 0.74 | 6.49 | 0.58 |
| | | ini = [-2,-2,2], tol = 2.0 | 0.0444 | 0.0110 | 1.0099 | 0.0555 | 13.85 | 1.06 | 10.08 | 0.93 |
| 3 | 5,000 | ini = [-0.5,-0.5,0.5], tol = 0.1 | 0.0040 | 0.0011 | 0.1126 | 0.0076 | 2.00 | 0.00 | 14.99 | 0.17 |
| | | ini = [-0.5,-0.5,0.5], tol = 0.5 | 0.0136 | 0.0021 | 0.3852 | 0.0138 | 2.00 | 0.00 | 14.94 | 0.09 |
| | | ini = [-0.5,-0.5,0.5], tol = 2.0 | 0.0378 | 0.0041 | 0.9941 | 0.0215 | 2.65 | 0.57 | 19.55 | 4.08 |
| | | ini = [-2,-2,2], tol = 0.1 | 0.0040 | 0.0011 | 0.1126 | 0.0076 | 5.00 | 0.00 | 36.21 | 0.80 |
| | | ini = [-2,-2,2], tol = 0.5 | 0.0136 | 0.0022 | 0.3852 | 0.0138 | 8.40 | 0.49 | 60.86 | 3.88 |
| | | ini = [-2,-2,2], tol = 2.0 | 0.0383 | 0.0040 | 0.9941 | 0.0215 | 13.55 | 0.50 | 93.44 | 6.03 |

Note: The reported values are averages and standard errors across 20 replications. $\alpha$ denotes the number of unimodal Gaussian distributions used when generating true parameters. |T| denotes the number of training markets. Computing time is measured in seconds. "ini" denotes the initial value of $\theta_0$. "tol" denotes the tolerance for error in market share. For a fair comparison, the number of taste clusters (M) is set to 1.

We then check the sensitivity of GLAM logit to the number of taste clusters ($M$), which is a key hyperparameter shaping the k-modal nonparametric distribution. Fig. 1 presents the distribution of true and estimated parameters in the multimodal scenario with $|T|$ in {500, 5,000} and $M$ in {1, 3, 5}. We set the initial values to [-0.5, -0.5, 0.5] and $tol$ to 0.1 to compare the results in the same context. The significant overlap between the true and estimated parameters in Fig.1 demonstrates the capability of GLAM logit to recover the taste heterogeneity of the true data generating process, which remains consistent across different values of $|T|$ and $M$. Fig. 1(a)-(b) illustrate that GLAM logit can recover multimodal parameter distributions even when the number of taste clusters is set to 1. This is because GLAM logit allows nonparametric estimation to identify the empirical parameter distribution that best fits the observed market share.

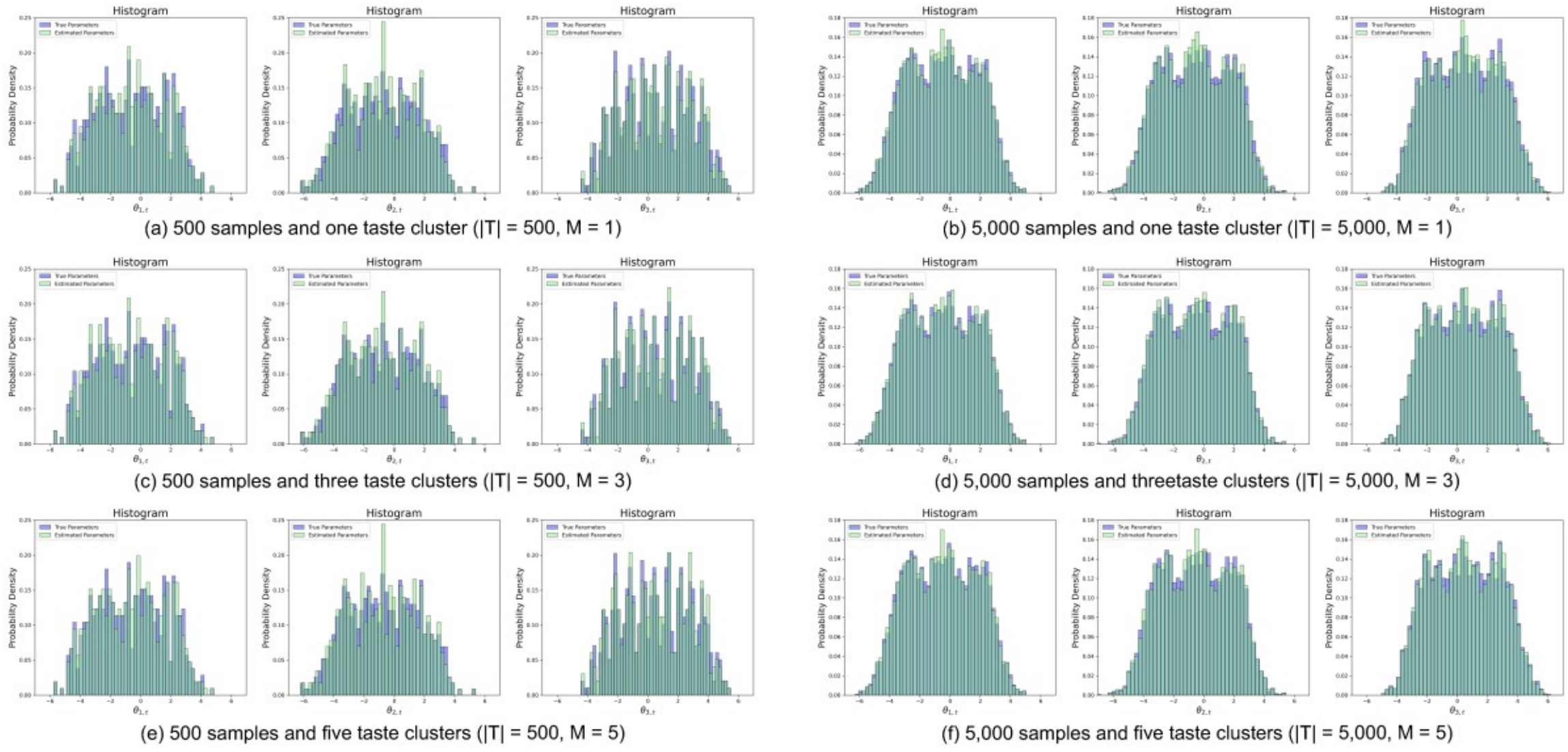

**Fig. 1.** Overlapping histograms of market-level true and estimated parameters in the multimodal scenario. In (a)-(f), x-axis is the value of parameters, y-axis is the probability density.

Fig. 2 further presents the distribution of taste parameters corresponding to the first attribute ($\theta_{1,t}$) in each taste cluster. The results indicate that when $M = 1$, the model approximates the overall distribution well, but it cannot identify the number of unimodal Gaussian distributions ($\alpha$) used for data generation. When the number of clusters ($M$) matches $\alpha$, the model more accurately captures the multimodal structure of the true parameter distribution, especially with a larger sample size (e.g., $|T| = 5{,}000$). However, when $M$ is set to a value larger than $\alpha$, a more complex multimodal distribution is captured, increasing the risk of overfitting.

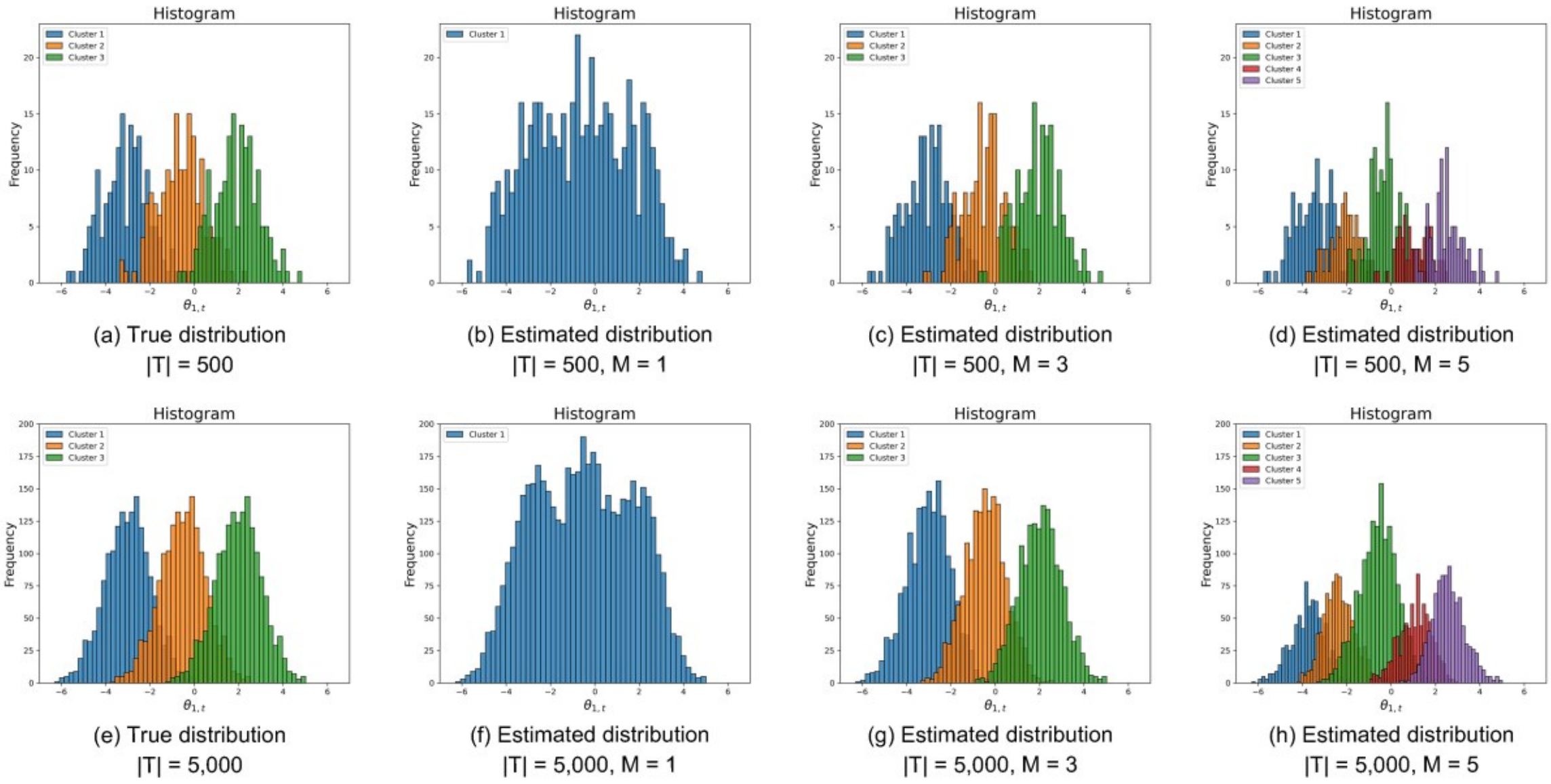

**Fig. 2.** Histograms of market-level true and estimated parameters per taste cluster corresponding to the first attribute. In (a)-(h), x-axis is the value of parameters, y-axis is the frequency.

We finally check the co-impacts of taste clusters and nearest neighbors on out-of-sample predictive accuracy. Table 3 demonstrates several important findings regarding the predictive performance of GLAM logit under different combinations of $M$ (number of clusters) and $\mathcal{K}$ (number of neighbors). First, GLAM logit achieves acceptable prediction accuracy overall, as shown by the consistent overall accuracy across different sample sizes. For example, when $\alpha = 1$ and $|T| = 500$, the highest overall accuracy reaches 82.60%, and remains stable at 82.82% when $|T| = 5{,}000$. This suggests that the KNN algorithm is effective when there are market-level attributes strongly related to taste parameters. Second, in experiments with smaller sample sizes and multimodal distributions (e.g., $\alpha = 3, |T| = 500$), larger $\mathcal{K}$ values lead to higher accuracy. For instance, the overall accuracy improves from 73.11% with $\mathcal{K} = 1$ to 76.90% with $\mathcal{K} = 5$ when $M = 3$. Finally, an interesting finding is that GLAM logit achieves the highest predictive accuracy when $M$ matches $\alpha$. For example, when $\alpha = 3$ and $|T| = 5{,}000$, the highest accuracy of 77.61% is observed with $M = 3$. These results suggest that it is possible to identify the best combination of $M$ and $\mathcal{K}$ based on out-of-sample predictive accuracy.

**Table 3**

Predictive performance of GLAM logit under different number of clusters ($M$) and nearest neighbors ($\mathcal{K}$)

| $\alpha$ | \|T\| | Method | Mean Absolute Error | | Overall Accuracy | | Adjusted R-Square | |
|---|---|---|---|---|---|---|---|---|
| | | | Mean | SE | Mean | SE | Mean | SE |
| 1 | 500 | GLAM logit (M = 1, $\mathcal{K}$ = 1) | 0.1112 | 0.0067 | 77.76% | 1.35% | 0.7114 | 0.0345 |
| | | GLAM logit (M = 1, $\mathcal{K}$ = 3) | 0.0911 | 0.0052 | 81.78% | 1.03% | 0.8030 | 0.0228 |
| | | **GLAM logit (M = 1, $\mathcal{K}$ = 5)** | 0.0870 | 0.0049 | 82.60% | 0.99% | 0.8225 | 0.0216 |
| | | GLAM logit (M = 3, $\mathcal{K}$ = 1) | 0.1280 | 0.0075 | 74.40% | 1.49% | 0.6397 | 0.0399 |
| | | GLAM logit (M = 3, $\mathcal{K}$ = 3) | 0.1052 | 0.0058 | 78.95% | 1.16% | 0.7530 | 0.0270 |
| | | GLAM logit (M = 3, $\mathcal{K}$ = 5) | 0.1005 | 0.0056 | 79.89% | 1.12% | 0.7774 | 0.0256 |
| 1 | 5,000 | GLAM logit (M = 1, $\mathcal{K}$ = 1) | 0.1093 | 0.0022 | 78.14% | 0.43% | 0.7227 | 0.0086 |
| | | GLAM logit (M = 1, $\mathcal{K}$ = 3) | 0.0904 | 0.0014 | 81.92% | 0.27% | 0.8103 | 0.0055 |
| | | **GLAM logit (M = 1, $\mathcal{K}$ = 5)** | 0.0859 | 0.0014 | 82.82% | 0.27% | 0.8297 | 0.0053 |
| | | GLAM logit (M = 3, $\mathcal{K}$ = 1) | 0.1257 | 0.0025 | 74.87% | 0.50% | 0.6548 | 0.0106 |
| | | GLAM logit (M = 3, $\mathcal{K}$ = 3) | 0.1044 | 0.0016 | 79.12% | 0.31% | 0.7619 | 0.0069 |
| | | GLAM logit (M = 3, $\mathcal{K}$ = 5) | 0.0992 | 0.0016 | 80.16% | 0.31% | 0.7827 | 0.0064 |
| 3 | 500 | GLAM logit (M = 1, $\mathcal{K}$ = 1) | 0.1687 | 0.0087 | 66.25% | 1.75% | 0.3209 | 0.0528 |
| | | GLAM logit (M = 1, $\mathcal{K}$ = 3) | 0.1536 | 0.0075 | 69.27% | 1.50% | 0.5092 | 0.0335 |
| | | GLAM logit (M = 1, $\mathcal{K}$ = 5) | 0.1492 | 0.0072 | 70.15% | 1.45% | 0.5365 | 0.0311 |
| | | GLAM logit (M = 3, $\mathcal{K}$ = 1) | 0.1345 | 0.0083 | 73.11% | 1.65% | 0.5734 | 0.0533 |
| | | GLAM logit (M = 3, $\mathcal{K}$ = 3) | 0.1187 | 0.0066 | 76.26% | 1.33% | 0.6469 | 0.0315 |
| | | **GLAM logit (M = 3, $\mathcal{K}$ = 5)** | 0.1155 | 0.0067 | 76.90% | 1.33% | 0.6656 | 0.0326 |
| 3 | 5,000 | GLAM logit (M = 1, $\mathcal{K}$ = 1) | 0.1628 | 0.0029 | 67.43% | 0.59% | 0.3516 | 0.0195 |
| | | GLAM logit (M = 1, $\mathcal{K}$ = 3) | 0.1483 | 0.0022 | 70.35% | 0.43% | 0.5294 | 0.0095 |
| | | GLAM logit (M = 1, $\mathcal{K}$ = 5) | 0.1452 | 0.0024 | 70.95% | 0.47% | 0.5473 | 0.0112 |
| | | GLAM logit (M = 3, $\mathcal{K}$ = 1) | 0.1290 | 0.0025 | 74.20% | 0.51% | 0.6065 | 0.0158 |
| | | **GLAM logit (M = 3, $\mathcal{K}$ = 3)** | 0.1120 | 0.0019 | 77.61% | 0.39% | 0.6765 | 0.0085 |
| | | GLAM logit (M = 3, $\mathcal{K}$ = 5) | 0.1142 | 0.0018 | 77.16% | 0.36% | 0.6658 | 0.0089 |

Note: The reported values are averages and standard errors across 20 replications. $\alpha$ denotes the number of unimodal Gaussian distributions used when generating true parameters. |T| denotes the number of markets. Computing time is measured in seconds. $\mathcal{K}$ denotes the number of nearest neighbors for prediction. For a fair comparison, the initial values (ini) are set to [-0.5,-0.5,0.5] and the tolerance level (tol) are set to 0.1.

## 5. Case study: travel mode choice modeling in New York State

### 5.1 Setup of experiments

*5.1.1 Data preparation*

The experiments are based on a synthetic population dataset provided by Replica Inc., which contains 53.55 million synthetic trips made by 19.53 million NYS residents on a typical Thursday in Fall 2019. The dataset was generated through a combination of census data, mobile phone data, economic activity data, and built environment data (Replica Inc., 2024). Information for each synthetic trip includes its origin, destination, travel mode, travel time, travel cost, and travelers' demographic attributes. Six travel modes are included: driving, public transit, on-demand auto, biking, walking, and carpool (trips made by several passengers in an auto vehicle).

There are two reasons for aggregating the synthetic dataset into market level. First, at individual level, the dataset only includes variables of the chosen modes; we do not know the travel time or cost of other alternatives. Second, individual trips are hard to validate but become more reliable when aggregated into census geo-units. According to the data quality report by Replica Inc. (2022), the largest error of demographic attributes is within 5% compared to census data, and the largest error of travel mode is within 10% compared to Census Transportation Planning Products (CTPP) data.

Hence, we aggregate the data based on population segments and trip origin-destination (OD) pairs. We consider four population segments: low-income, not-low-income, senior, and student population. Firstly, we identify the student population still in schools, colleges, and universities. We then identify the senior population whose age is over 65. To differentiate the low-income and not-low-income populations, we refer to U.S. Federal Poverty Guidelines[2]. Moreover, we use census block group units for spatial aggregation. Trips belonging to the same block group-level OD pair are averaged to retrieve the market shares and variables of the six modes. Finally, 53.55 million individual trips are aggregated into 120,740 unique markets, with each market represented by an agent. Fig. 3 visualizes these agents in New York State. Table 4 provides a summary of variables in the aggregate dataset for modeling. We use 80% of the total agents for training the model, and the remaining 20% are used for testing.

[2] https://aspe.hhs.gov/topics/poverty-economic-mobility/poverty-guidelines

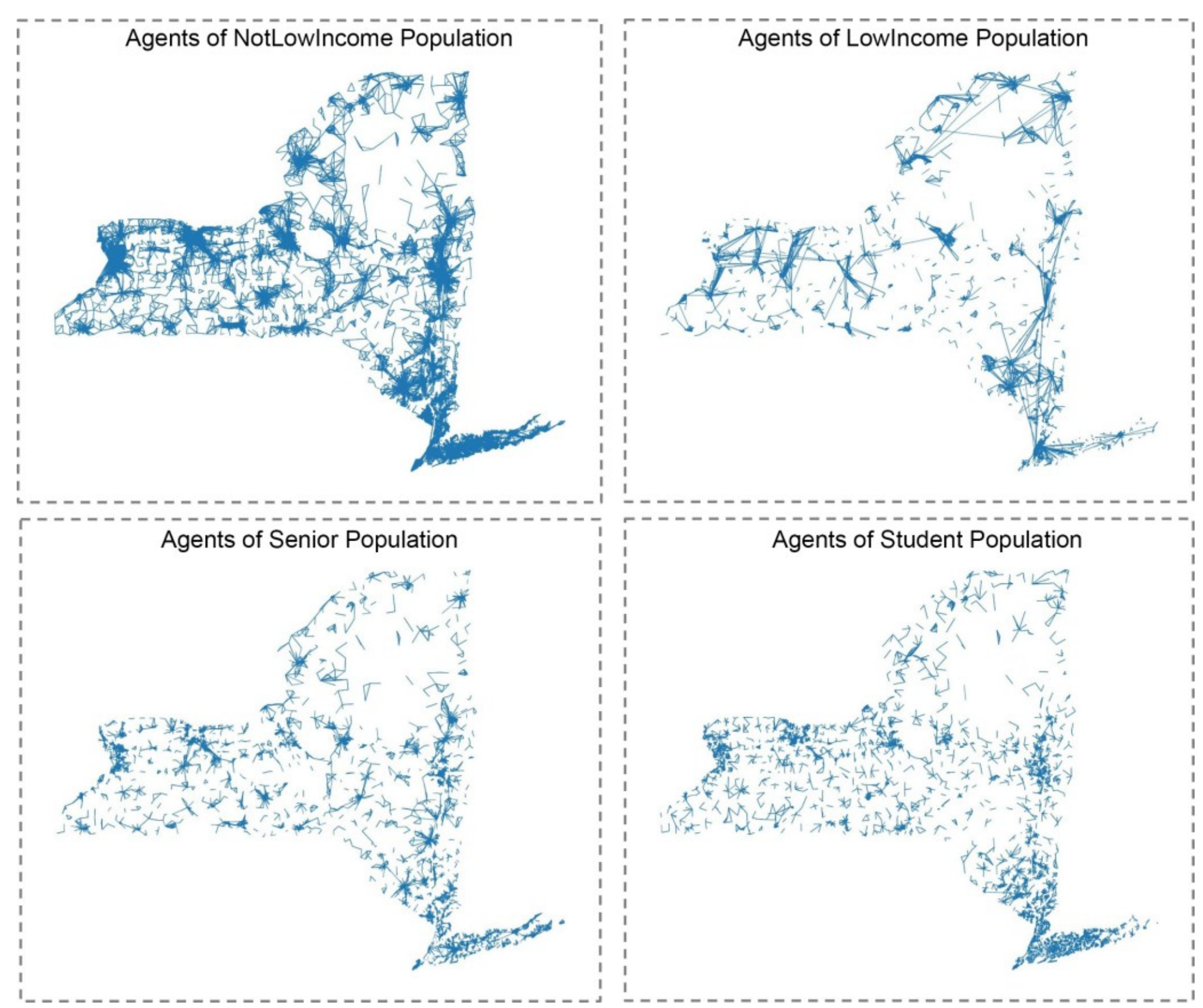

**Fig. 3.** Agents per population segment in New York State

**Table 4**
Summary of variables in the aggregate dataset for modeling.

| Variable | Count | Mean | Std. | Min. | 25% | 50% | 75% | Max. |
|---|---|---|---|---|---|---|---|---|
| **Driving** | | | | | | | | |
| Travel time (min) | 120,740 | 6.51 | 6.48 | 0.06 | 2.46 | 4.62 | 8.28 | 200.78 |
| Travel cost ($) | 120,740 | 0.67 | 1.12 | 0.01 | 0.11 | 0.24 | 0.72 | 9.54 |
| Market share (%) | 120,740 | 56.44% | 30.99% | 0% | 28.36% | 68.97% | 82.00% | 100% |
| **Public transit** | | | | | | | | |
| Access time (min) | 120,740 | 6.40 | 6.49 | 0.06 | 2.30 | 4.47 | 8.42 | 220.57 |
| Egress time (min) | 120,740 | 7.16 | 7.21 | 0.07 | 2.60 | 5.01 | 9.43 | 234.22 |
| In-vehicle time (min) | 120,740 | 16.31 | 23.89 | 2.07 | 5.43 | 10.57 | 20.17 | 834.00 |
| Transfers | 120,740 | 0.44 | 0.35 | 0 | 0.11 | 0.42 | 0.72 | 6.00 |
| Travel cost ($) | 120,740 | 1.55 | 0.61 | 0.69 | 1.38 | 1.38 | 1.38 | 2.75 |
| Market share (%) | 120,740 | 3.38% | 13.14% | 0% | 0% | 0% | 0% | 100% |
| **On-demand auto** | | | | | | | | |
| Travel time (min) | 120,740 | 6.44 | 6.44 | 0.06 | 2.38 | 4.50 | 8.14 | 200.54 |
| Travel cost ($) | 120,740 | 10.24 | 9.99 | 0.69 | 4.40 | 7.21 | 12.14 | 532.843 |
| Market share (%) | 120,740 | 1.11% | 2.80% | 0% | 0% | 0% | 1.28% | 75.00% |
| **Biking** | | | | | | | | |
| Travel time (min) | 120,740 | 14.67 | 14.61 | 0.11 | 5.20 | 10.50 | 19.25 | 457.97 |
| Market share (%) | 120,740 | 0.67% | 1.90% | 0% | 0% | 0% | 0.36% | 62.50% |
| **Walking** | | | | | | | | |
| Travel time (min) | 120,740 | 28.27 | 28.04 | 0.25 | 10.84 | 21.42 | 34.93 | 892.51 |
| Market share (%) | 120,740 | 22.02% | 28.70% | 0% | 0% | 8.75% | 33.94% | 100% |
| **Carpool** | | | | | | | | |
| Travel time (min) | 120,740 | 6.59 | 6.56 | 0.06 | 2.50 | 4.72 | 8.38 | 340.12 |
| Travel cost ($) | 120,740 | 7.44 | 6.89 | 0.01 | 3.73 | 5.50 | 8.89 | 232.57 |
| Market share (%) | 120,740 | 16.38% | 17.97% | 0% | 6.17% | 11.11% | 17.74% | 100% |

Note: The unit of time is transformed to hours when building the model.

*5.1.2 Model specification for GLAM logit*

In line with existing mode choice studies, we consider a series of variables including travel time, travel cost, number of transfers, and mode specific constants (we assume waiting time is included into travel time due to data availability). The systematic utilities of the sixed modes can be written as Eqs. (37)-(42).

$$V_{driving,t} = \theta_{tt_{auto},t} Time_{driving,t} + \theta_{cost,t} Cost_{driving,t} + \theta_{asc_{driving},t}, \quad \forall t \in T \tag{37}$$

$$V_{transit,t} = \theta_{at_{transit},t} AT_{transit,t} + \theta_{et_{transit},t} ET_{transit,t} + \theta_{ivt_{transit},t} IVT_{transit,t} + \theta_{nt_{transit},t} NT_{transit,t} + \theta_{cost,t} Cost_{transit,t} + \theta_{asc_{transit},t}, \quad \forall t \in T \tag{38}$$

$$V_{on-demand,t} = \theta_{tt_{auto},t} Time_{on-demand,t} + \theta_{cost,t} Cost_{on-demand,t} + \theta_{asc_{on-demand},t}, \quad \forall t \in T \tag{39}$$

$$V_{biking,t} = \theta_{tt_{non-auto},t} Time_{biking,t} + \theta_{asc_{biking},t}, \forall t \in T \tag{40}$$

$$V_{walking,t} = \theta_{tt_{non-auto},t} Time_{walking,t} + \theta_{asc_{walking},t}, \quad \forall t \in T \tag{41}$$

$$V_{carpool,t} = \theta_{tt_{auto},t} Time_{carpool,t} + \theta_{cost,t} Cost_{carpool,t}, \quad \forall t \in T \tag{42}$$

where $Time_{driving,t}$, $IVT_{transit,t}$, $Time_{on-demand,t}$, $Time_{biking,t}$, $Time_{walking,t}$, $Time_{carpool,t}$ are the travel time of six modes in agent $t$; $AT_{transit,t}$, $ET_{transit,t}$, $NT_{transit,t}$ are the access time, egress time, and number of transfers of public transit; $Cost_{driving,t}$, $Cost_{transit,t}$, $Cost_{on-demand,t}$, $Cost_{carpool,t}$ are the travel cost of modes except biking and carpool. $\theta_{tt_{auto},t}$, $\theta_{ivt_{transit},t}$, $\theta_{at_{transit},t}$, $\theta_{et_{transit},t}$, $\theta_{nt_{transit},t}$, $\theta_{tt_{non-auto},t}$, $\theta_{cost,t}$, $\theta_{asc_{driving},t}$, $\theta_{asc_{transit},t}$, $\theta_{asc_{on-demand},t}$, $\theta_{asc_{biking},t}$, $\theta_{asc_{walking},t}$ are 12 parameters per agent $t$ to be estimated. Since the parameters of travel time and cost variables are expected to be negative, we set the upper boundary of these parameters to zero.

Following Krueger et al. (2023)'s work, we treat the travel cost of on-demand mode as an endogenous variable. We create instrumental variables using the approach adopted by Fosgerau et al. (2024) and run instrumental regression. First, we group the six modes across two dimensions – auto mode (driving, on-demand auto, carpool) versus non-auto mode (transit, walking, biking), and mode with waiting time (transit, on-demand auto, carpool) versus mode without waiting time (driving, biking, walking). For each dimension, travel time variables of other modes in the same group are averaged. Since we have three travel time variables related to auto travel time, transit in-vehicle time, and non-auto travel time, we create six instrumental variables for the two dimensions. Finally, we run instrumental regression on on-demand travel cost ($Cost_{on-demand,t}$in Eq. (39)) and replace it with the one predicted by the regression model. This would result in an unbiased estimation of $\theta_{cost,t}$.

For the hyperparameter selection, we set the initial values to a vector of zeros as we do not know the true values of taste parameters. We set the tolerance level to 0.1 according to the results in our simulation study. The number of taste clusters ($M$) and nearest neighbors ($\mathcal{K}$) are selected based on the out-of-sample performance defined in Section 4.2.2. We use 80% of the data for training and the remaining 20% for testing. For each test market, we first pick out training markets that belong to the same population segment. The KNN algorithm is then applied using the geolocations of trip origins and destinations as input features to retrieve the taste parameters. Table 5 shows the out-of-sample predictive accuracy of GLAM logit models with different combinations of $M$ and nearest neighbors $\mathcal{K}$. The combination of $M = 2$ and $\mathcal{K} = 3$ results in the highest performance in this case study, although a number of other combinations achieve very close accuracy. Therefore, we set $M = 2$ and $\mathcal{K} = 3$ for the following experiments.

**Table 5**
Out-of-sample predictive accuracy under different numbers ($M$) of clusters and nearest neighbors ($\mathcal{K}$).

| | $M = 1$ | $\boldsymbol{M = 2}$ | $M = 3$ | $M = 4$ | $M = 5$ |
|---|---|---|---|---|---|
| Mean absolute error (out-of-sample) | | | | | |
| $\mathcal{K} = 1$ | 0.0381 | 0.0361 | 0.0360 | 0.0371 | 0.0375 |
| $\mathcal{K} = 2$ | 0.0347 | 0.0321 | 0.0325 | 0.0338 | 0.0333 |
| $\boldsymbol{\mathcal{K} = 3}$ | 0.0338 | **0.0302** | 0.0318 | 0.0319 | 0.0327 |
| $\mathcal{K} = 4$ | 0.0324 | 0.0305 | 0.0307 | 0.0316 | 0.0318 |
| $\mathcal{K} = 5$ | 0.0322 | 0.0308 | 0.0306 | 0.0312 | 0.0314 |
| Overall accuracy (out-of-sample) | | | | | |
| $\mathcal{K} = 1$ | 77.12% | 78.61% | 78.42% | 77.86% | 77.63% |
| $\mathcal{K} = 2$ | 79.43% | 80.13% | 80.36% | 79.45% | 78.70% |
| $\boldsymbol{\mathcal{K} = 3}$ | 79.84% | **81.78%** | 81.11% | 79.69% | 80.51% |
| $\mathcal{K} = 4$ | 80.13% | 81.69% | 81.67% | 81.25% | 81.06% |
| $\mathcal{K} = 5$ | 80.79% | 81.40% | 81.58% | 81.52% | 81.20% |
| Adjusted R-square (out-of-sample) | | | | | |
| $\mathcal{K} = 1$ | 0.7612 | 0.7746 | 0.7718 | 0.7671 | 0.7652 |
| $\mathcal{K} = 2$ | 0.7890 | 0.7954 | 0.7968 | 0.7890 | 0.7863 |
| $\boldsymbol{\mathcal{K} = 3}$ | 0.7911 | **0.8059** | 0.8047 | 0.7979 | 0.7979 |
| $\mathcal{K} = 4$ | 0.7963 | 0.8053 | 0.8050 | 0.8032 | 0.8030 |
| $\mathcal{K} = 5$ | 0.7987 | 0.8041 | 0.8052 | 0.8049 | 0.8047 |

*5.1.3 Model specification for benchmarks*

An essential part of our experiments is to benchmark GLAM logit against current market-level models. Based on the literature review, we build multinomial logit (MNL), nested logit (NL), inverse product differentiation logit (IPDL), and the BLP model as benchmarks. We treat the carpool mode as the reference level or outside alternative ($s_{0t} = s_{carpool,t}$), and all variables in Eqs. (37)-(42) are transformed to values relative to carpool. The price endogeneity in Eq. (39) is addressed using the same IV approach in the estimation of benchmark models. Since the markets are defined by four population segments related to income and age, the fixed effects of these socio-demographics have been included in GLAM logit and all benchmark models.

Following, Huo et al. (2024)'s work, we estimate MNL, NL, and IPDL by solving a linear instrumental regression on the logarithm form of market share ratio ($\ln\left(\frac{s_{jt}}{s_{0t}}\right)$). For instance, the ratios of driving market share to carpool market share in MNL, NL, IPDL are defined in Eqs. (43)-(45).

$$\ln\left(\frac{s_{driving,t}}{s_{carpool,t}}\right) = \theta_{tt_{auto}} Time_{driving,t}^{carpool} + \theta_{cost} Cost_{driving,t}^{carpool} + \theta_{asc_{driving}}, \quad \forall t \in T \tag{43}$$

$$\ln\left(\frac{s_{driving,t}}{s_{carpool,t}}\right) = \theta_{tt_{auto}} Time_{driving,t}^{carpool} + \theta_{cost} Cost_{driving,t}^{carpool} + \rho_{auto} \ln\left(\frac{s_{driving,t}}{\sum_{j\in J_{auto}} s_{jt}}\right) + \theta_{asc_{driving}}, \quad \forall t \in T \tag{44}$$

$$\ln\left(\frac{s_{driving,t}}{s_{carpool,t}}\right) = \theta_{tt_{auto}} Time_{driving,t}^{carpool} + \theta_{cost} Cost_{driving,t}^{carpool} + \rho_{auto} \ln\left(\frac{s_{driving,t}}{\sum_{j\in J_{auto}} s_{jt}}\right) + \rho_{waiting} \ln\left(\frac{s_{driving,t}}{\sum_{j\in J_{waiting}} s_{jt}}\right) + \theta_{asc_{driving}}, \quad \forall t \in T \tag{45}$$

where $Time_{driving,t}^{carpool}$, $Cost_{driving,t}^{carpool}$ are driving time and cost relative to carpool; $\theta_{tt_{auto}}$, $\theta_{cost}$, $\theta_{asc_{driving}}$ are parameters to be estimated (referring to $\alpha$ and $\beta$ in market-level models); $J_{auto} = \{driving, ondemand\}$ is the set of auto modes, $J_{waiting} = \{transit, ondemand\}$ is the set of modes with waiting time. We specify nests based on auto mode for the NL model, and we consider both auto mode and waiting time for product segmentation in the IPDL model. These models are estimated using the AER package in R.

As for the BLP model, we set the normally distribution on parameters of auto travel time, transit in-vehicle time, non-auto travel time, and travel cost. The two-step estimation of the BLP model is conducted using the PyBLP package in Python developed by Conlon and Gortmaker (2020).

*5.1.4 Elasticity estimates*

In addition to the predictive accuracy defined in Section 4.2.2, we compute two measures to compare the elasticity estimates of GLAM logit and benchmark models. Elasticity is a good metric to identify substitution patterns. For a fair comparison, we increase the travel cost (price) of driving, transit, on-demand auto by 1%, predict mode shares with our models, and calculate the percentage change of mode shares to obtain direct- and cross-price price elasticity, as shown in Eq. (46).

$$e_j^{p_{j^*}} = \frac{1}{|T|} \sum_{t\in T} \left(\frac{\Delta \hat{s}_{jt}/\hat{s}_{jt}}{\Delta p_{j^*t}/p_{j^*t}}\right), \quad \forall j, j^* \in J \tag{46}$$

where $e_j^{p_{j^*}}$ denotes mode $j$'s elasticity regarding mode $j^*$'s travel cost; $\Delta p_{jt}/p_{jt}$ is the percentage change of mode $j^*$'s travel cost that is 1% in our experiments; $\Delta \hat{s}_{jt}/\hat{s}_{jt}$ is the percentage change of $j$'s mode share predicted by our models. $j = j^*$ results in direct elasticity and $j \neq j^*$ results in cross elasticity.

The diversion ratio is a metric that helps identify both substitution and complementarity (Huo et al., 2024). In our experiments, the diversion ratio from mode $j^*$ to $j$ is defined as the negative of the ratio of trips that shifted from mode $j^*$ to $j$ and all trips that shifted from mode $j^*$ due to a 1% increase in the travel time of mode $j^*$, as shown in Eq. (47).

$$D_{j^*j} = \frac{1}{|T|} \sum_{t\in T} \left(-\frac{\Delta \hat{s}_{jt}}{\Delta \hat{s}_{j^*t}}\right), \quad \forall j, j^* \in J \tag{47}$$

where $\Delta\hat{s}_{jt}$, $\Delta\hat{s}_{j^*t}$ are changes in market shares due to a 1% increase in the travel time of mode $j^*$. A positive diversion ratio implies that mode $j^*$ and $j$ are substitutes, while a negative diversion ratio implies complementarity. Moreover, we have $\sum_{j\in J, j\neq j^*} D_{j^*j} = 1, \forall j^* \in J$ and $D_{j^*j} = -1$ if $j^* = j$.

## 5.2 Model results

This section presents the results of GLAM logit and benchmark models. The experiments were performed on a local machine equipped with an Intel Core i7-10875H CPU and 32GB of RAM. The stopping criteria $\epsilon$ was set to $10^{-3}$. GLAM logit was estimated using the Gurobi package in Python, while the AER package in R was used for MNL, NL, and IPDL estimations. For the BLP model, the PyBLP package in Python was used. We present the model results from three aspects: (1) basic statistics; (2) predictive accuracy; and (3) elasticity estimates.

### *5.2.1 Basic statistics*

Table 6 summarizes the mean values, standards error, and significant levels of models built with training data, from which we can compare the GLAM logit model to benchmark models under the same context. Standard errors in GLAM logit are bootstrapped using 30 resamples. Several interesting points are found.

(1) The parameters estimated by MNL, NL, IPDL, BLP, and GLAM logit show great consistency in signs: the parameters of travel time and travel cost have negative signs (besides transit in-vehicle time in MNL and NL), the constants of driving and walking have positive signs, and the constants of public transit, on-demand auto, and biking have negative signs. These results are aligned with our empirical knowledge.

(2) All parameters in our models are significant at 0.1% level, which is partially due to our large sample size. The significance of nest parameters indicates that the mode segmentation in NL and IPDL is appropriate. The significance of control variable parameter indicates that endogeneity correction in the GLAM logit model is necessary.

(3) GLAM logit took 2h 31mins to converge given 96,592 training agents, which is much longer than MNL (35 s), NL (46 s), and IPDL (55 s). However, such an estimation time is still acceptable compared to market-level models with random parameters, since the BLP model took 37 h 14 min to converge with only four random parameters (and the BLP model failed to converge when we set all parameters to be random).

**Table 6**
Parameter estimates of GLAM logit and benchmark models (each entry represents the average value of one estimated parameter, and the number in the parenthesis is the standard error).

| | **MNL** | **NL** | **IPDL** | **BLP** | | **GLAM logit** | |
|---|---|---|---|---|---|---|---|
| | **Mean** | **Mean** | **Mean** | **Mean** | **SD** | **Mean** | **SD** |
| **Travel time and cost** | | | | | | | |
| Auto travel time | -6.21*** | -3.79*** | -5.21*** | -4.64*** | 1.09*** | -2.27*** | 0.46*** |
| ($\theta_{tt_{auto},t}$) | (0.13) | (0.21) | (0.09) | (0.04) | (0.16) | (0.01) | (3E-03) |
| Transit in-vehicle | 0.55*** | 0.21*** | -0.32*** | -4.10*** | 1.54*** | -2.07*** | 1.24*** |
| time ($\theta_{ivt_{transit},t}$) | (0.02) | (0.02) | (0.02) | (0.42) | (0.19) | (0.01) | (8E-03) |
| Transit access time | -4.70*** | -4.81*** | -4.22*** | -9.76*** | | -1.01*** | 0.59*** |
| ($\theta_{at_{transit},t}$) | (0.21) | (0.27) | (0.17) | (0.06) | | (0.01) | (6E-03) |
| Transit egress time | -5.32*** | -5.79*** | -4.01*** | -6.12*** | | -1.12*** | 0.69*** |
| ($\theta_{et_{transit},t}$) | (0.19) | (0.25) | (0.15) | (0.19) | | (0.01) | (7E-03) |
| Number of transfers | -1.47*** | -0.99*** | -1.01*** | -0.94*** | | -3.20*** | 0.89*** |
| ($\theta_{nt_{transit},t}$) | (0.02) | (0.02) | (0.17) | (0.03) | | (0.01) | (4E-03) |
| Non-vehicle travel | -5.20*** | -4.36*** | -2.58*** | -3.53*** | 0.06*** | -4.29*** | 2.40*** |
| time ($\theta_{tt_{non-auto},t}$) | (0.02) | (0.08) | (0.02) | (0.08) | (0.17) | (0.02) | (0.01) |
| Trip cost | -0.01*** | -0.04*** | -0.07*** | -1.07*** | 1.15*** | -0.32*** | 2.66*** |
| ($\theta_{cost,t}$) | (2E-03) | (2E-03) | (8E-03) | (1E-03) | (0.04) | (5E-03) | (4E-03) |
| **Mode specific constant** | | | | | | | |
| Driving constant | 0.55*** | 0.45*** | 0.10*** | 1.12*** | | 0.32*** | 1.51*** |
| ($\theta_{asc_{driving},t}$) | (0.02) | (0.03) | (0.01) | (0.01) | | (9E-03) | (7E-03) |
| Transit constant | -3.41*** | -2.07*** | -3.24*** | -0.04*** | | -2.74*** | 1.21*** |
| ($\theta_{asc_{transit},t}$) | (0.02) | (0.04) | (0.01) | (0.11) | | (0.01) | (0.01) |
| On demand constant | -4.38*** | -2.79*** | -4.06*** | -4.99*** | | -2.42*** | 1.52*** |
| ($\theta_{asc_{on-demand},t}$) | (0.01) | (0.03) | (0.01) | (4E-03) | | (0.02) | (0.03) |
| Biking constant | -3.93*** | -2.64*** | -0.91*** | -3.19*** | | -4.07*** | 1.22*** |
| ($\theta_{asc_{biking},t}$) | (0.01) | (0.02) | (0.01) | (0.03) | | (0.01) | (0.02) |
| Walking constant | 0.86*** | 1.14*** | 0.29*** | 1.37*** | | 0.60*** | 1.60*** |
| ($\theta_{asc_{walking},t}$) | (9E-03) | (0.03) | (2E-03) | (0.06) | | (0.01) | (0.02) |
| **Nest parameter** | | | | | | | |
| $\ln\left(\frac{s_{jt}}{\sum_{j' \in J_{auto}} s_{j't}}\right)$ | | 0.40***<br>(3E-03) | 0.29***<br>(2E-03) | | | | |
| $\ln\left(\frac{s_{jt}}{\sum_{j' \in J_{waiting}} s_{j't}}\right)$ | | | 0.59***<br>(2E-03) | | | | |
| **Meta information** | | | | | | | |
| Instrumental variables (IVs) | Yes | Yes | Yes | Yes | | Yes | |
| # Observations | 96,592 | 96,592 | 96,592 | 96,592 | | 96,592 | |
| Estimation time | 35 s | 46 s | 55 s | 37 h 14 mins | | 2 h 31 mins | |

Note: ***p-value<0.001, **p-value<0.01, *p-value<0.05

*5.2.2 Prediction accuracy*

Table 7 shows the prediction accuracy of GLAM logit and benchmark models. The in-sample prediction accuracy reflects the model goodness-of-fit. IPDL outperforms MNL and NL, which can be attributed to its flexibility in product segmentation. The BLP model outperforms IPDL by allowing four parameters to be normally distributed, which validates the existence of taste heterogeneity. Thus, the BLP model has the highest in-sample prediction accuracy among benchmarks. Our findings differ from Huo et al. (2024)'s work, in which IPDL performed better than the BLP model with an automobile dataset. This might be because their dataset contains 624 products and 31 markets, while our dataset includes 6 modes and 96,592 markets/agents. To this end, the importance of capturing taste heterogeneity is higher than identifying product segmentation when the number of markets is much larger than the number of products in a market-level model.

The in-sample performance of GLAM logit is considerably superior compared to the BLP model. GLAM logit reduces the mean absolute error from 0.0305 to 0.0109, improving the overall accuracy from 78.70% to 96.42%, and improving the adjusted R-square from 0.8060 to 0.9744. This is because GLAM logit specifies agent-specific parameters, leading to a flexible non-parametric distribution fitting to the observed data.

The out-of-sample accuracy reflects the reliability of model predictions with new datasets and indicates the extent of overfitting. The out-of-sample predictive performance of all models generally dropped, but GLAM logit still maintains superior performance, with the difference in overall out-of-sample accuracy of GLAM logit and the BLP model being 16.48%. Considering that GLAM logit estimates a unique set of parameters for each agent, such overfitting is acceptable because the relative differences in out-of-sample performance of GLAM and benchmarks is similar to that of in-sample performance.

Two characteristics of GLAM logit help address overfitting issues: (1) Individual trips are aggregated into markets, which makes GLAM logit more robust than individual-level models with similar estimation approaches (Ren & Chow, 2022) ; (2) The KNN algorithm further reduces the risk of overfitting.

**Table 7**
Predictive accuracy of GLAM logit and benchmark models.

| | Mean absolute error | Overall accuracy (%) | Adjusted R-square |
|---|---|---|---|
| **In-sample predictive accuracy** | | | |
| MNL | 0.0863 | 56.45% | 0.6682 |
| NL | 0.0790 | 58.72% | 0.7077 |
| IPDL | 0.0432 | 71.28% | 0.7734 |
| BLP | 0.0305 | 78.70% | 0.8060 |
| GLAM logit | 0.0109 | 96.42% | 0.9744 |
| **Out-of-sample predictive accuracy** | | | |
| MNL | 0.0925 | 54.97% | 0.6143 |
| NL | 0.0767 | 56.21% | 0.6671 |
| IPDL | 0.0541 | 61.39% | 0.7193 |
| BLP | 0.0458 | 65.30% | 0.7377 |
| GLAM logit | 0.0302 | 81.78% | 0.8059 |

Note: In-sample accuracy are calculated using 80% of the markets. Out-of-sample accuracy are calculated using the rest 20%.

*5.2.3 Elasticity estimates*

We further compare the elasticity estimates of GLAM logit to IPDL and the BLP model (two models with the highest performance in benchmarks). Aggregated direct- and cross-price elasticities over six modes are presented in Table 8. The magnitude of direct elasticity in IPDL is larger compared to GLAM logit, which are larger than those of BLP, but in general their trends are similar: (1) The scale of direct-price elasticities is larger than cross-price elasticities, indicating that modes are more sensitive to their own travel cost compared to travel cost of other modes. (2) For direct-price elasticity, the three modes ranked by the sensitivity to their own travel cost are on-demand auto (-0.144 in IPDL, -0.297 in BLP, -0.198 in GLAM logit), public transit (-0.0465 in IPDL, -0.0127 in BLP, -0.0230 in GLAM logit), and driving (-0.00708 in IPDL, -0.00171 in BLP, -0.00375 in GLAM logit). (3) For cross-price elasticity, no negative value is found among the six modes and IPDL cross-price elasticity estimates are quite close to those of GLAM logit.

GLAM logit offers flexibility to set an upper boundary on parameter estimates to ensure that the parameters for time and cost remain negative. This prevents each market from exhibiting unreasonable elasticities, such as an increase in transit travel time leading to higher ridership.

**Table 8**
Comparison of price elasticity estimates in IPDL, BLP, and GLAM logit.

| | Direct | Cross | | | | | |
|---|---|---|---|---|---|---|---|
| | | Driving | Transit | On-demand | Biking | Walking | Carpool |
| **IPDL** | | | | | | | |
| Driving | -7.08E-03 | -- | 4.51E-03 | 3.15E-03 | 5.13E-03 | 6.35E-03 | 2.57E-03 |
| Transit | -4.65E-02 | 6.51E-04 | -- | 4.32E-03 | 2.65E-03 | 2.19E-03 | 1.00E-03 |
| On-demand | -1.44E-01 | 2.00E-03 | 4.39E-03 | -- | 1.72E-03 | 1.65E-03 | 1.59E-03 |
| **BLP** | | | | | | | |
| Driving | -1.71E-03 | -- | 8.09E-03 | 5.51E-03 | 4.77E-03 | 2.94E-03 | 3.93E-03 |
| Transit | -1.27E-02 | 3.24E-04 | -- | 8.51E-04 | 6.32E-04 | 2.73E-04 | 3.25E-04 |
| On-demand | -2.97E-01 | 3.30E-03 | 4.29E-03 | -- | 2.54E-03 | 1.88E-03 | 2.49E-03 |
| **GLAM logit** | | | | | | | |
| Driving | -3.75E-03 | -- | 4.91E-03 | 4.31E-03 | 4.59E-04 | 5.92E-03 | 3.33E-03 |
| Transit | -2.30E-02 | 4.07E-04 | -- | 3.01E-03 | 1.93E-03 | 2.05E-03 | 5.41E-04 |
| On-demand | -1.98E-01 | 2.47E-03 | 4.01E-03 | -- | 2.12E-03 | 1.60E-03 | 2.02E-03 |

Note: We only consider the price of driving, public transit, and on-demand auto since biking and walking are free and carpool is set as the outside alternative in benchmark models.

Fig. 4 visualizes the diversion ratios in IPDL, BLP, and GLAM logit. The diversion ratio measures the proportion of trips that switch from one mode to another when there is a 1% increase in travel time for the original mode. The diagonal values are all negative, indicating that an increase in travel time for a given mode results in a decrease in demand for that mode, as expected. Moreover, higher diversion ratios (cells in yellow) in these models indicate that a majority of trips shifted from the primary mode to driving, walking, and carpool. However, the off-diagonal values are all positive, suggesting that no complementarity is found even in the IPDL model.

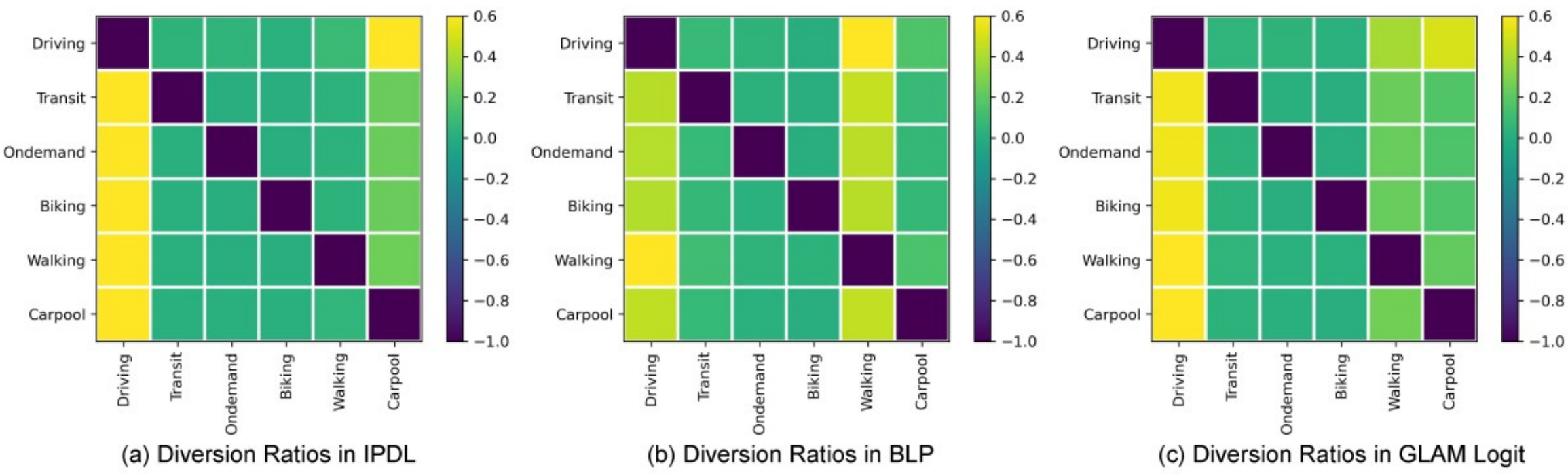

**Fig. 4.** Comparison of diversion ratios in IPDL, BLP, and GLAM logit. In (a)-(c), x-axis denotes modes to calculate diversion ratios ($\boldsymbol{j}$), y-axis denotes modes with a 1% increase of travel time ($\boldsymbol{j}^*$).

## 5.3 Empirical parameter distribution and its applications

This section aims to showcase the empirical taste heterogeneity captured by GLAM logit and how these results can be applied to further analysis.

### *5.3.1 Empirical distribution of taste parameters*

Fig. 5(a) and (d) show the mean values of parameters in each iteration, from which we can see the GLAM logit model converged at the 27th iteration of Algorithm 1. Fig. 5(b)-(c) and (e)-(f) present the parameter distribution of two taste clusters after the final iteration, revealing that the empirical parameter distribution does not resemble traditionally considered distributions (e.g., Gaussian and uniform). Twelve parameters in two clusters are generally unimodal and can be divided into two categories: (1) highly concentrated parameters with non-zero means, such as transit access and egress time ($\theta_{at_{transit},t}$, $\theta_{et_{transit},t}$) in cluster 1, indicating homogeneous tastes among agents; and (2) evenly distributed parameters with non-zero means, such as non-auto travel time ($\theta_{tt_{non-auto},t}$) in cluster 2 and driving constant ($\theta_{asc_{driving},t}$) in cluster 1, indicating heterogeneous tastes among agents. Moreover, the two taste clusters are different from each other. Compared to cluster 1, cluster 2 has larger negative values of non-auto travel time ($\theta_{tt_{non-auto},t}$) and transit in-vehicle time ($\theta_{ivt_{transit},t}$) but a larger positive value of driving constant ($\theta_{asc_{driving},t}$). To this end, cluster 2 can be labeled as "driving lovers" and cluster 1 can be labeled as "non-driving lovers" or "others".

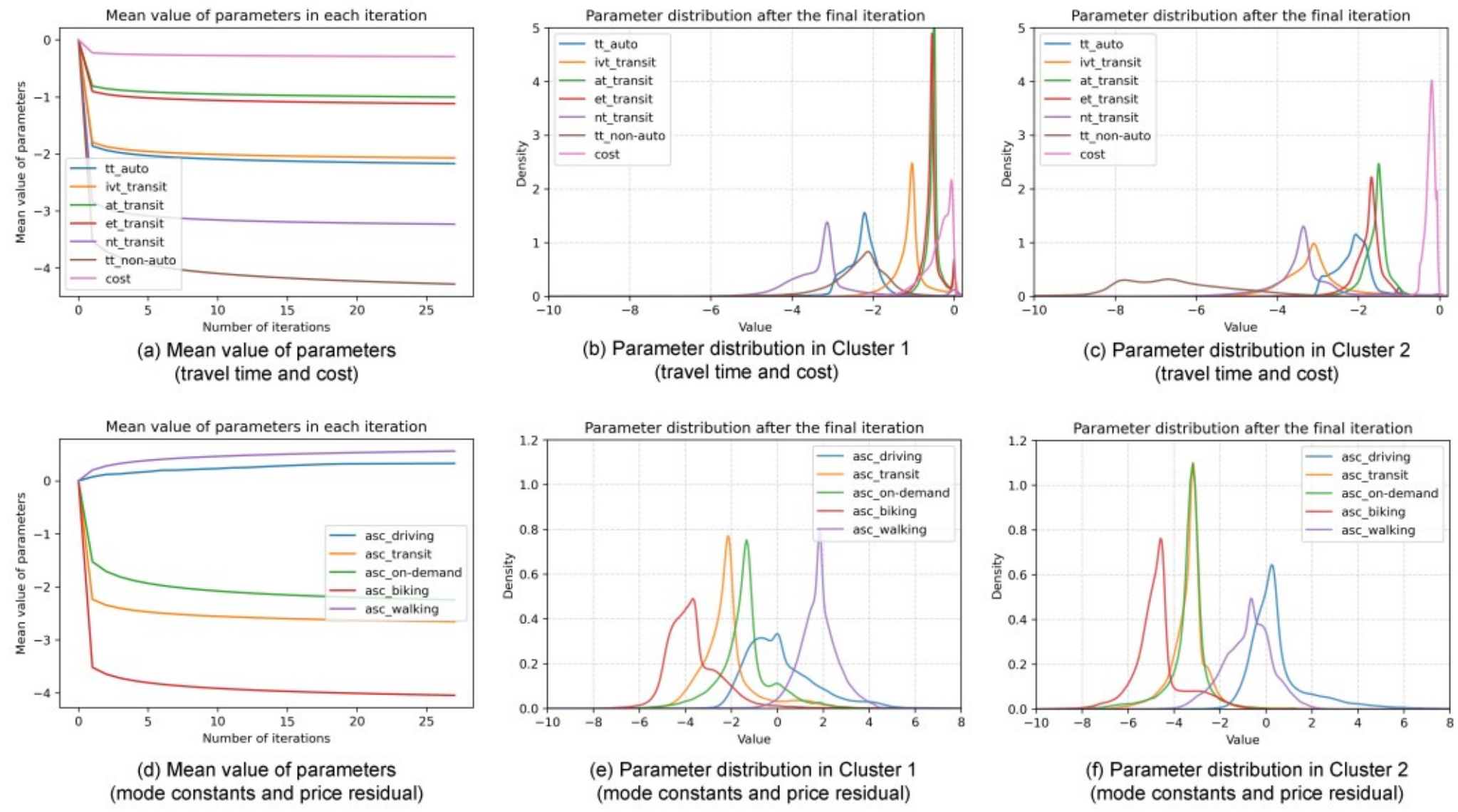

**Fig. 5.** Mean values and parameter distributions. In (a), (d), x-axis is the number of iterations, y-axis is parameter mean value. In (b)-(c), (e)-(f), x-axis is the value of estimated parameters, y-axis is the probability density.

Since GLAM logit provides market-specific taste parameters, we can further explore the taste heterogeneity among population segments and regions. Table 9 lists the average value-of-time (VOT) of four population segments in New York State and New York City. VOT is measured as the marginal rate of substitution between travel time and cost. The results are consistent with existing studies and our empirical knowledge (Chow et al., 2020; Lam & Small, 2001). On the one hand, the average VOT in New York City is generally higher than in New York State. On the other hand, the not-low-income populations have the highest VOT (\$18.11/hour in NYS and \$28.53/hour in NYC) while low-income populations have the lowest VOT (\$8.92/hour in NYS and \$9.62/hour in NYC). It is worth emphasizing that benchmark models cannot capture these differences unless we build a separate model for each segment or interaction effects are hand-crafted in the utility equation.

**Table 9**
Value of time (VOT) of different population segments (each entry represents the average VOT, and the number in the parenthesis is the standard deviation).

| | **VOT in NY State (\$/hour)** | **VOT in NYC (\$/hour)** |
|---|---|---|
| Not-low-income Population | 18.11 (7.66) | 28.53 (17.29) |
| Low-income Population | 8.92 (3.68) | 9.62 (6.01) |
| Senior Population | 12.07 (4.58) | 13.23 (6.37) |
| Student Population | 9.94 (4.56) | 11.65 (6.63) |

The agent-specific parameters allow us to plot taste heterogeneity in space. Fig. 6 shows the spatial distribution of VOT (value-of-time) in New York State and New York City. In New York State, the VOT in New York City, Albany, Buffalo, Syracuse, Rochester, and Ithaca is noticeably higher than in other areas. Among these cities, NYC has the highest VOT. Within NYC, trips related to Manhattan and trips heading to JFK airport have relatively higher VOT, while trips in Staten Island have relatively lower VOT. These details, uniquely captured by our GLAM logit model, can serve as valuable references for the operating strategies of statewide transportation services.

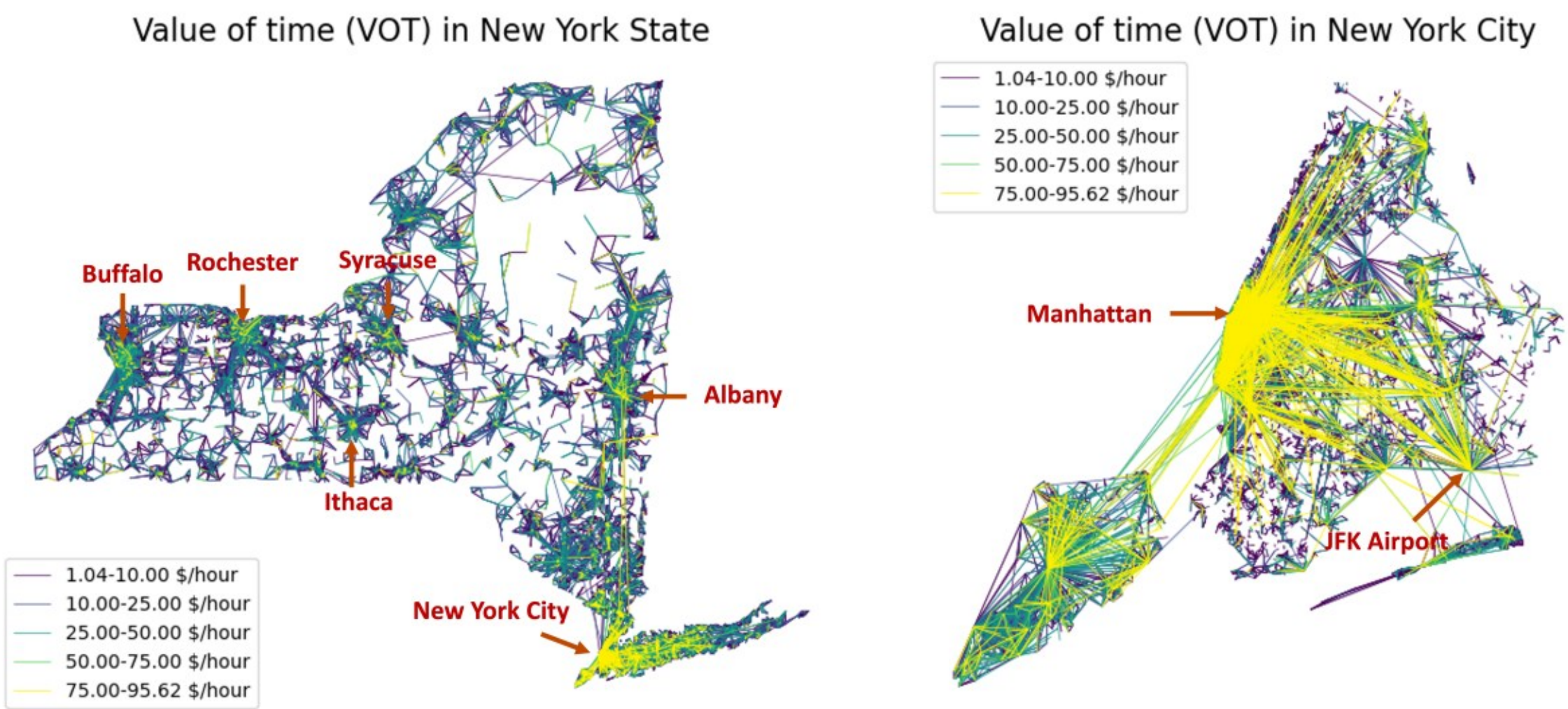

**Fig. 6.** Distribution of Value of time (VOT) in NY State and NYC.

*5.3.2 Application 1: travelers' response to congestion pricing*

Congestion pricing in New York City took effect on January 5th, 2024, which involves charging passenger and small commercial vehicles a \$9 toll to enter Manhattan south of 60th Street during peak hours (MTA, 2025). Fig. 7 (a) shows a map highlighting the boundaries of the Congestion Relief Zone. In our case study, each market refers to a number of trips made by a population segment from one census block group to another. Hence, the estimated taste parameters can be applied to evaluating how sensitive travelers' mode choices are to the congestion toll.

To do this, we calculate the compensating variation (CV) for the driving mode using parameters estimated by the GLAM logit model. This measures the amount of money travelers would need to be compensated per trip to maintain their utility level if driving were to become unavailable. According to Chipman and John (1980)'s work, CV for driving mode in market $t$ ($CV_{driving,t}$) can be written as Eq. (48).

$$CV_{driving,t} = \frac{1}{\theta_{cost,t}}\left(\ln\left(\sum\nolimits_{j\in J^-} e^{V_{jt}}\right) - \ln\left(\sum\nolimits_{j\in J} e^{V_{jt}}\right)\right), \quad \forall t \in T \tag{48}$$

where $\theta_{cost,t}$ is the parameter of travel cost estimated by GLAM logit, $V_{jt}$ is the systematic utility of mode $j$ in agent $t$, $J$ is the original choice set with six modes, and $J^-$ is the choice set without driving mode. A CV that falls below \$9 suggests that a traveler would be willing to trade away the auto mode option, i.e. shift to another mode, if charged a congestion fee of \$9/trip. Conversely, a CV greater than \$9 means the traveler values the auto mode more than the fee that they would pay and would not be willing to switch.

Two comparisons are made. First, a comparison is made between Not-low-income with all the other segments (Low-income, Senior, Student) to show whether there's a significance difference in elasticity that may warrant subsidies for low income, senior, and student segments. A second comparison is made between residents of NYC entering the zone versus everyone else in NYS entering the zone. Note that this study only examines NYS starting and ending trips, so it doesn't include trips originating from NJ, CT, or other states. It also does not consider other choices like changing departure times or destination. Moreover, since our dataset only includes trip OD pairs, we ignore specific trips and hours and treat all trips that end in the Congestion Relief Zone as impacted trips, resulting in 59,645 trips/day in total.

Fig. 7 (b)-(c) present cumulative density functions (CDFs) of driving CV for these trips. Several interesting points are found: (1) The \$9 toll exceeds the CV for ~60% of the total trips entering the congestion zone, indicating that a majority of travelers would be willing to drop driving option and consider alternative modes (at least during peak hours) due to the toll; (2) The CV for not-low-income population is generally higher than for other populations, which means it is more challenging to shift not-low-income population from driving to other modes, or vice versa, only ~\$4 is needed to nudge the same proportion of other populations to switch mode as \$9 for not-low-income; (3) Only about 10% of trips starting outside of NYC have a CV lower than \$9, indicating that these trips are less likely to shift modes due to the congestion toll (i.e. a higher percentage of them will end up paying the toll). This might be due to the inconvenience of other modes for trips from upstate NY or Long Island to NYC.

To the best of our knowledge, this is one of the first logit-based analysis of the elasticity of all NYS travelers' mode choice to the MTA congestion pricing toll. Future research can also include optimization of the toll price to maximize consumer surplus and revenue.

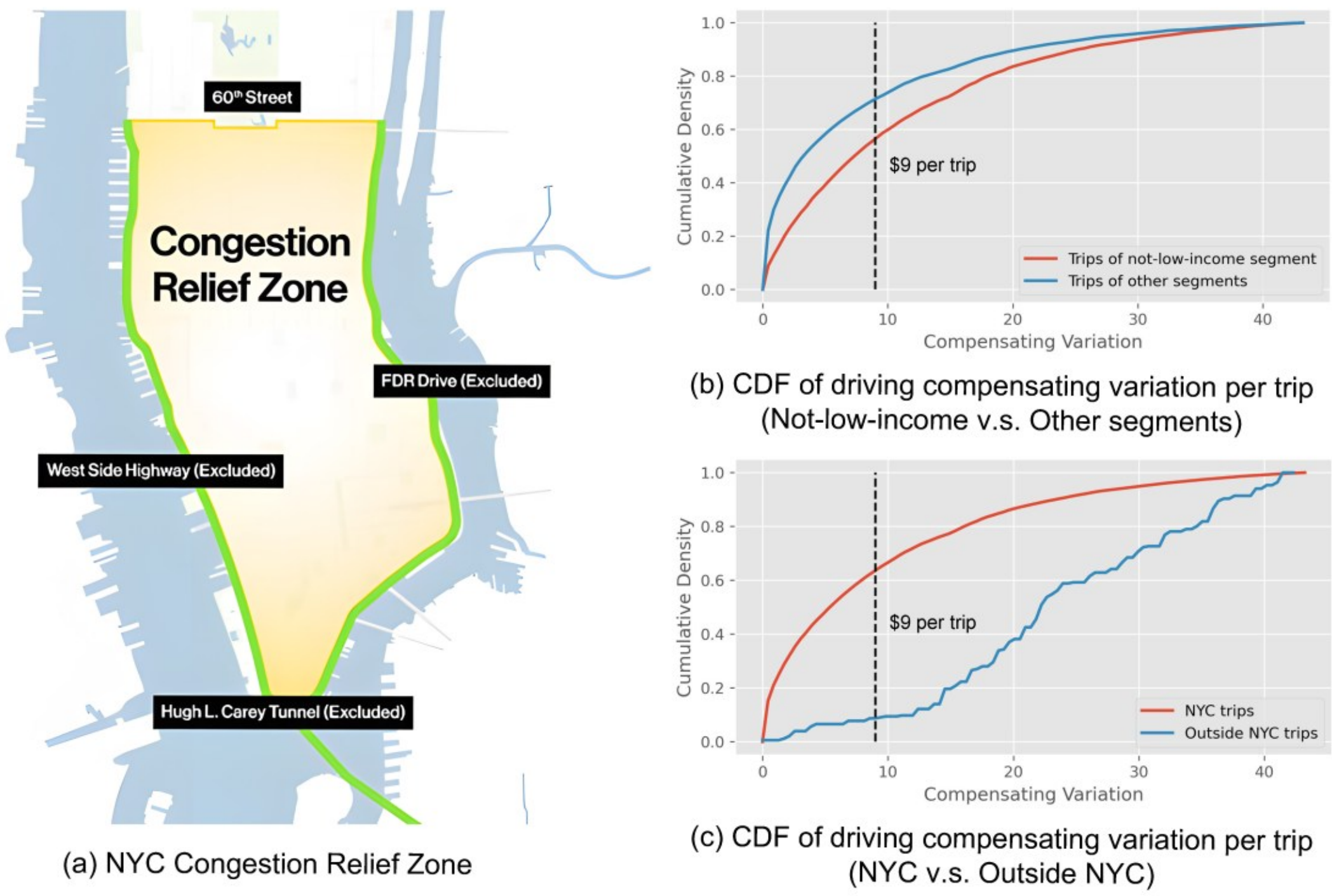

(a) NYC Congestion Relief Zone

(b) CDF of driving compensating variation per trip (Not-low-income v.s. Other segments)

(c) CDF of driving compensating variation per trip (NYC v.s. Outside NYC)

**Fig. 7.** Cumulative density function of driving compensating variation (CV) for trips end in NYC Congestion Relief Zone. (a) is a screenshot from https://congestionreliefzone.mta.info/tolling. In (b)-(c), x-axis is the value of CV for driving mode, y-axis is the cumulative density.

*5.3.3 Application 2: transit fare discounting to increase ridership*

Since the utility function of each market is fully specified in GLAM logit, its estimation results can be directly incorporated into optimization models for revenue management and system design. In this section, we provide a simple demo to showcase how GLAM logit helps link the demand and supply sides efficiently.

Let us assume that the state government provides a 50% fare discount to encourage public transit ridership. The discount would be issued county by county with different fare revenue loss per county. Given a budget level of acceptable fare revenue loss, the task is to maximize the increase in ridership by selecting a number of counties to apply the discount. This scenario can be formulated as a binary programming (BP) problem, where the ridership before and after the transit fare discount is calculated using our GLAM logit model. The BP problem is formulated as shown in Eqs. (49)-(55).

$$\max_{y_i, x_t} \sum_{t \in T} \hat{s}_{transit,t} d_t \tag{49}$$

subject to:

$$\hat{s}_{transit,t} = \hat{s}_{transit,t}^{dis} x_t + \hat{s}_{transit,t}^{non-dis}(1 - x_t), \qquad \forall t \in T \tag{50}$$

$$\sum_{i \in I} y_i \leq O \tag{51}$$

$$\sum_{t \in T} 0.5 * c_{transit,t} x_t \leq B, \qquad \forall t \in T \tag{52}$$

$$\sum_{t \in T_i} x_t = |T_i| y_i, \qquad \forall\, i \in I \tag{53}$$

$$y_i \in \{0,1\}, \qquad \forall\, i \in I \tag{54}$$

$$x_t \in \{0,1\}, \qquad \forall t \in T \tag{55}$$

where $y_i$ is a binary variable indicating whether county $i$ is selected to apply the discount, $x_t$ is a binary variable indicating whether agent $t$ is impacted by the discount, $I$ is the set of all counties in NY state, and $T_i$ is the set of all agents in county $i$. $d_t$ and $c_{transit,t}$ are the total travel demand (trips/day) and transit fare (\$/trip) for agent $t$, which can be obtained from the synthetic population data. $\hat{s}_{transit,t}^{dis}$, $\hat{s}_{transit,t}^{non-dis}$ are market shares with and without the discount that can be predicted by GLAM logit in advance. Eq. (50) ensures that the final predicted market share ($\hat{s}_{transit,t}$) equals $\hat{s}_{transit,t}^{dis}$ if $x_t = 1$ and $\hat{s}_{transit,t}^{non-dis}$ if $x_t = 0$. $O$ determines the maximum number of counties with the fare discount, and Eq. (51) ensures that the number of selected counties is no more than $O$. $B$ determines the budget level, and Eq. (52) ensures that the revenue loss per day due to the discount is no more than $B$. Eq. (53) ensures that all agents in county $i$ will have the discount once the county is selected. All the equations are in closed form, making the BP problem efficient to solve.

In our instance, the BP problem contains 62 binary decision variables for counties ($y_i$) and 120,740 binary decision variables for agents ($x_t$). We solve it with the Gurobi package in Python, which only took 12 seconds to get the optimal solution. Table 10 summarizes the optimization results given $O = 10$ and $B$ equal to \$5,000, \$50,000, and \$500,000, respectively. Fig. 8 visualizes the selected counties and agents in space. When $B = \$5{,}000$, transit ridership would increase by 1,896 trips/day, and the optimal solution suggests selecting counties outside metropolitan areas. When $B = \$50{,}000$, transit ridership would increase by 9,402 trips/day, and the optimal solution suggests selecting counties in major cities. When $B = \$500{,}000$, transit ridership would increase by 13,201 trips/day, and the selected counties include the whole NYC. Additionally, the maximum revenue loss is \$196,123/day under our settings. Since the aim of this section is to illustrate a further application of GLAM logit, some of our assumptions might not be realistic. For a more elaborate case, we refer interested readers to Ren et al. (2024)'s work.

**Table 10**
A summary of transit ridership and revenue in the ten selected counties.

| | **Total ridership** | **Total revenue** | **Change of ridership** | **Change of revenue** |
|---|---|---|---|---|
| $B$ = \$5,000 | 194,234 trips/day | \$444,358/day | +1,896 trips/day | -\$4,995/day |
| $B$ = \$50,000 | 201,763 trips/day | \$403,813/day | +9,402 trips/day | -\$45,326/day |
| $B$ = \$500,000 | 205,812 trips/day | \$255,013/day | +13,201 trips/day | -\$196,123/day |

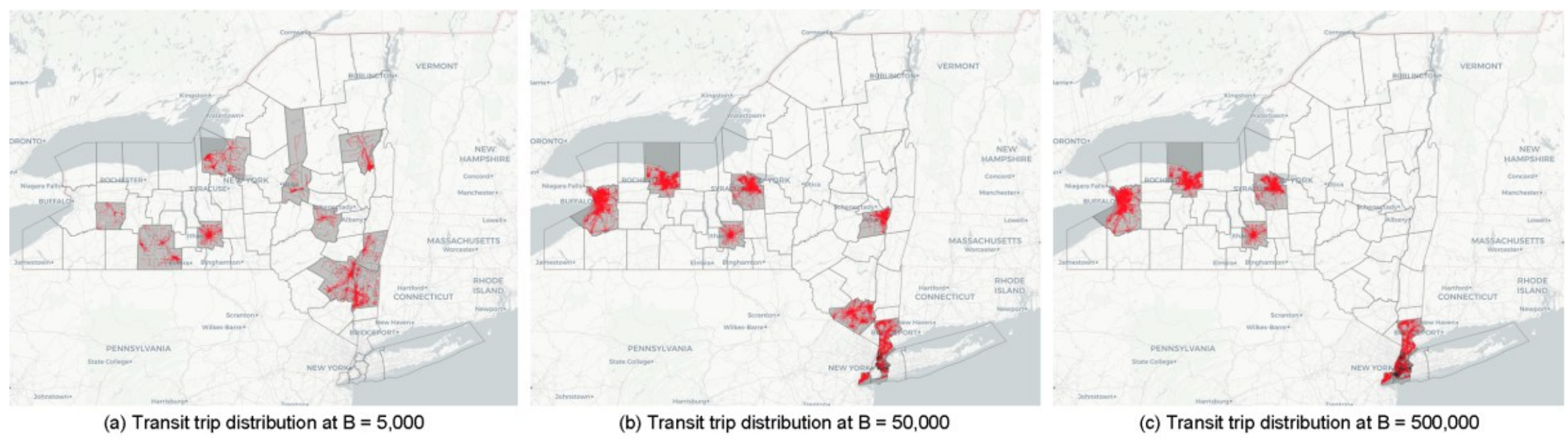

(a) Transit trip distribution at B = 5,000 (b) Transit trip distribution at B = 50,000 (c) Transit trip distribution at B = 500,000

**Fig. 8.** Visualization of the optimization results

## 6. Conclusion

Though large-scale ICT datasets contain individual mobility profiles, it is more feasible and reliable to build choice models with aggregate data to address privacy concerns, resolve issues associated with unobserved choice sets, and reduce measurement errors in location-based datasets. This study presents a group-level agent-based mixed (GLAM) logit model, which estimates agent-specific parameters by solving a multiagent inverse utility maximization (MIUM) problem with taste clusters. This method is designed to overcome the limitations of existing market-level models in capturing taste heterogeneity while ensuring scalability and computational tractability.

The simulation study evaluates the performance of GLAM logit under various hyperparameter configurations. Results show that GLAM logit is more stable when the sample size reaches 5,000, and it achieves global convergence with a small $tol$. The model performs best when the number of taste clusters ($M$) aligns with the complexity of the true parameter distribution ($\alpha$). Predictive accuracy also remains consistent across sample sizes, with higher number of nearest neighbors ($\mathcal{K}$) improving performance in multimodal scenarios with smaller sample size. The best combination of $M$ and $\mathcal{K}$ can be found based on out-of-sample predictive accuracy.

The application of the GLAM logit model in a mode choice case study for New York State demonstrates its superior in-sample and out-of-sample performance compared to benchmark models, including MNL, NL, IPDL, and BLP. The GLAM logit model achieves a significant improvements in overall accuracy (over 15% compared to BLP), highlighting its robustness and predictive power. Meanwhile, the GLAM logit model provides direct and cross-price elasticity estimates similar to the benchmark models. Furthermore, the market-level parameters in GLAM logit allow for further analyses of value-of-time (VOT) and taste heterogeneity across different population segments and regions. This level of detail can inform targeted transportation policies and optimize service delivery.

Despite the advantages outlined above, there remain many research opportunities and challenges to be addressed. First, our current results are based on a dataset with a small number of products (modes). Case studies with larger choice sets (e.g., route choice, destination choice/accessibility, etc.) are required to further validate the model's applicability. In such cases, it is possible to combine the strengths of GLAM logit and IPDL. For instance, the product differentiation components in Eq. (13) can serve as additional attributes in GLAM logit, allowing for generic or market-level parameters estimation. Second, the proposed approach can only capture static preferences. Incorporating temporal dynamics into the model could provide valuable insights into how taste heterogeneity evolves over time. Additionally, allowing individuals to vary within a market is another meaningful direction. This could be achieved by modifying the formulation of the MIUM problem at the beginning, such as employing bi-level estimation to capture both intra- and inter-market heterogeneity. Last but not least, improving the computational efficiency of GLAM logit is another direction of our future study, which would allow our model to handle even larger datasets and more complex choice scenarios.

## Acknowledgments

Funding support from C2SMARTER and NYU's Summer Undergraduate Research Program are appreciated. Data shared by Replica Inc. are gratefully acknowledged. Prateek Bansal acknowledges support from the presidential young professiorship (PYP) grant.